\documentclass[prX,aps,twocolumn,superscriptaddress,showpacs,longbibliography]{revtex4-1}

\usepackage{amssymb}
\usepackage{epsfig}
\usepackage{subfigure}
\usepackage{graphicx}
\usepackage{textcomp}
\usepackage{float}
\usepackage[normalem]{ulem}
\usepackage{color}


\expandafter\let\csname equation*\endcsname\relax
\expandafter\let\csname endequation*\endcsname\relax
\usepackage{amsmath}

\begin{document}

\title{Safe Leads and Lead Changes in Competitive Team Sports}
\author{A. Clauset}
\affiliation{Department of Computer Science, University of Colorado, Boulder, CO 80309, USA}
\affiliation{BioFrontiers Institute, University of Colorado, Boulder, CO 80305, USA}
\affiliation{Santa Fe Institute, 1399 Hyde Park Road, Santa Fe, NM 87501, USA}
\author{M. Kogan}
\affiliation{Department of Computer Science, University of Colorado, Boulder, CO 80309, USA}
\author{S. Redner}
\affiliation{Santa Fe Institute, 1399 Hyde Park Road, Santa Fe, NM 87501, USA}
\affiliation{Center for Polymer Studies and Department of Physics, Boston University, Boston, MA 02215, USA}

\begin{abstract}
  We investigate the time evolution of lead changes within individual games
  of competitive team sports.  Exploiting ideas from the theory of random
  walks, the number of lead changes within a single game follows a Gaussian
  distribution.  We show that the probability that the last lead change and
  the time of the largest lead size are governed by the same arcsine law, a
  bimodal distribution that diverges at the start and at the end of the game.
  We also determine the probability that a given lead is ``safe'' as a
  function of its size $L$ and game time $t$.  Our predictions generally
  agree with comprehensive data on more than 1.25 million scoring events in
  roughly 40,000 games across four professional or semi-professional team
  sports, and are more accurate than popular heuristics currently used in
  sports analytics.  
\end{abstract}

\maketitle

\section{Introduction}

Competitive team sports, including, for example, American football, soccer,
basketball and hockey, serve as model systems for social competition, a
connection that continues to foster intense popular interest.  This passion
stems, in part, from the apparently paradoxical nature of these sports. On
one hand, events within each game are unpredictable, suggesting that chance
plays an important role.  On the other hand, the athletes are highly skilled
and trained, suggesting that differences in ability are fundamental.  This
tension between luck and skill is part of what makes these games exciting for
spectators and it also contributes to sports being an exemplar for
quantitative modeling, prediction and human
decision-making~\cite{mosteller1997lessons,palacios2003professionals,ayton:fischer:2004,albert2005anthology},
and for understanding broad aspects of social competition and
cooperation~\cite{barney:1986,michael:chen:2005,romer:2006,johnson:2006,berger:2011,radicchi:2012}.

In a competitive team sport, the two teams vie to produce events (``goals'')
that increase their score, and the team with the higher score at the end of
the game is the winner. (This structure is different from individual sports
like running, swimming and golf, or judged sports, like figure skating,
diving, and dressage.)~ We denote by $X(t)$ the instantaneous difference in
the team scores.  By viewing game scoring dynamics as a time series, many
properties of these competitions may be quantitatively
studied~\cite{reed:hughes:2006,galla:farmer:2013}.  Past work has
investigated, for example, the timing of scoring
events~\cite{thomas:2007,everson2008composite,heuer:etal:2010,buttrey2011estimating,yaari:eisenman:2011,gabel:redner:2012,merritt:clauset:2014},
long-range correlations in scoring~\cite{ribeiro2012anomalous}, the role of
timeouts~\cite{saavedra:etal:2012}, streaks and ``momentum'' in
scoring~\cite{gilovich1985hot,vergin:2000,sire:redner:2009,arkes2011finally,yaari:eisenman:2011,yaari:david:2012},
and the impact of spatial positioning and playing field
design~\cite{bourbousson:seve:mcgarry:2012,merritt:clauset:2013}.

In this paper, we theoretically and empirically investigate a simple yet
decisive characteristic of individual games: the times in a game when the
lead changes.  A lead change occurs whenever the score difference $X(t)$
returns to 0.  Part of the reason for focusing on lead changes is that these
are the points in a game that are often the most exciting.  Although we are
interested in lead-change dynamics for all sports, we first develop our
mathematical results and compare them to data drawn from professional
basketball, where the agreement between theory and data is the most
compelling.  We then examine data for three other major competitive American
team sports:\ college and professional football, and professional hockey, and
we provide some commentary as to their differences and similarities.

Across these sports, we find that many of their statistical properties are
explained by modeling the evolution of the lead $X$ as a simple random walk.
More strikingly, seemingly unrelated properties of lead statistics,
specifically, the distribution of the times $t$: (i) for which one team is
leading $\mathcal{O}(t)$, (ii) for the last lead change $\mathcal{L}(t)$, and
(iii) when the maximal lead occurs $\mathcal{M}(t)$, are all described by the
same celebrated arcsine law~\cite{levy:1939,levy:1948,morters:peres:2010}:
\begin{equation}
\label{arcsine}
\mathcal{O}(t)=\mathcal{L}(t)=\mathcal{M}(t)= \frac{1}{\pi} \frac{1}{\sqrt{t(T-t)}} \enspace ,
\end{equation}
for a game that lasts a time $T$.  These three results are, respectively, the
\emph{first}, \emph{second}, and \emph{third} arcsine laws.

Our analysis is based on a comprehensive data set of all points scored in
league games over multiple consecutive seasons each in the National
Basketball Association (abbreviated NBA henceforth), all divisions of NCAA
college football (CFB), the National Football League (NFL), and the National
Hockey League (NHL)~\footnote{Data provided by STATS LLC, copyright
  2015.}. These data cover 40,747 individual games and comprise 1,306,515
individual scoring events, making it one of the largest sports data sets
studied.
Each scoring event is annotated with the game clock time $t$ of the event, its point
value, and the team scoring the event.  For simplicity, we ignore events
outside of regulation time (i.e., overtime).  We also combine the point
values of events with the same clock time (such as a successful foul shot
immediately after a regular score in basketball).  Table~\ref{tab:data}
summarizes these data and related facts for each sport.

\begin{table*}[t!]
\begin{center}
\begin{tabular}{ll|ccc|ccccc}
&  &  Num.\  & Num.\ scoring & Duration & Mean events  & Mean pts.\  & Persistence & Mean num.\ & Frac.\ with no\\\ 
Sport~~ & Seasons~~~~~ &  games &  events & $T$ (sec) & per game $N$ & per event $s$ &  $p$ & lead changes $\mathcal{N}$ & lead changes \\ \hline
NBA & 2002--2010 & 11,744 & 1,098,747 & 2880 & 93.56 & 2.07 & 0.360 & 9.37 & 0.063\\
CFB & 2000--2009  & 14,586 & \hspace{0.7em}123,448 & 3600 & \hspace{0.5em}8.46 & 5.98 & 0.507 & 1.23 & 0.428\\
NFL & 2000--2009  & \hspace{0.5em}2,654 & \hspace{1.2em}20,561 & 3600 & \hspace{0.5em}7.75 & 5.40 & 0.457 & 1.43 & 0.348\\
NHL & 2000--2009  & 11,763 &  \hspace{1.2em}63,759 & 3600 & \hspace{0.5em}5.42 & 1.00 & --- & 1.02 & 0.361
\end{tabular}
\end{center}
\caption{Summary of the empirical game data for the team sports considered in this study, based
 on regular-season games and scoring events within regulation time. }
\label{tab:data}
\end{table*}

\subsection*{Basketball as a model competitive system}

\begin{figure}[H]
  \centerline{\includegraphics[width=0.4\textwidth]{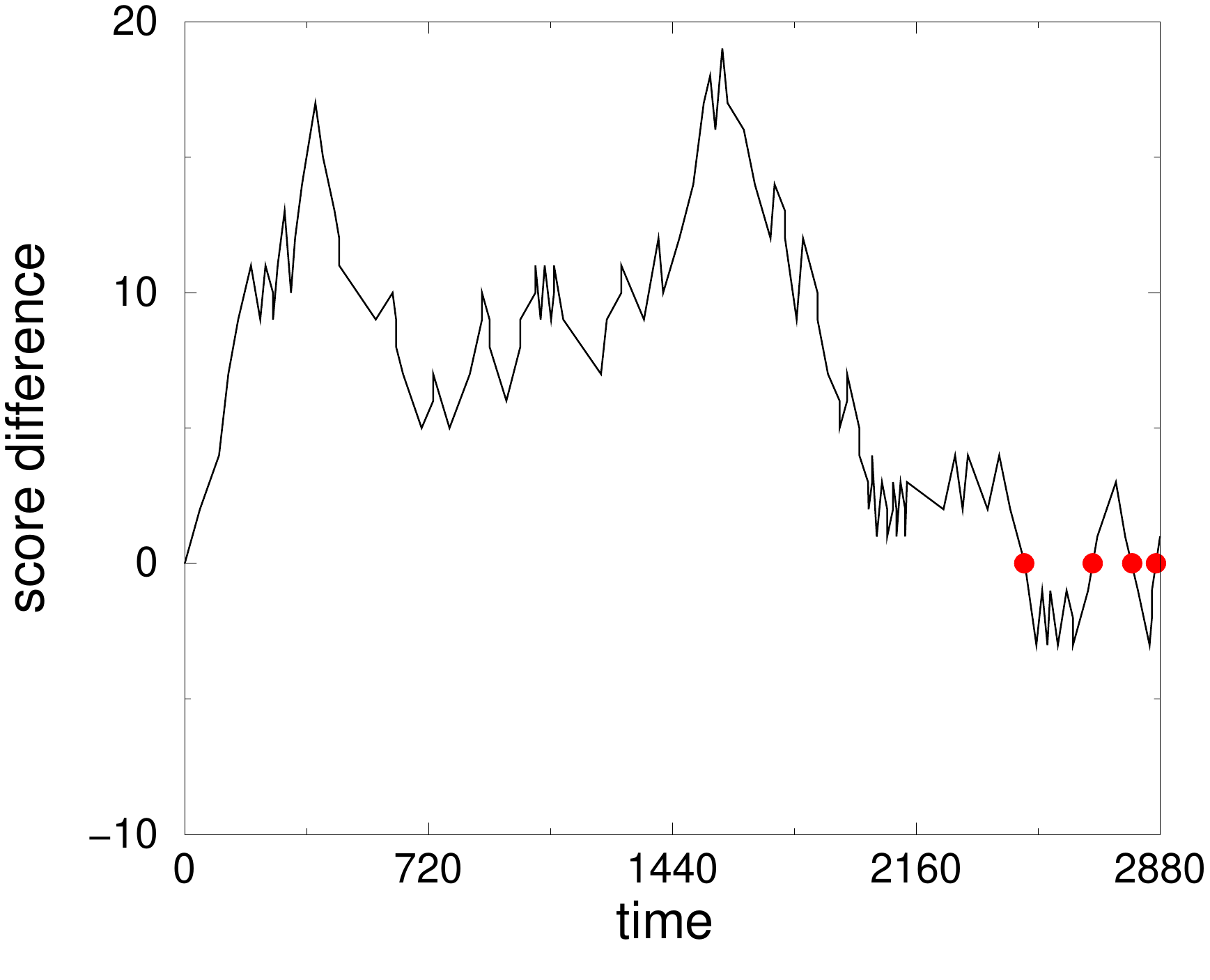}}
  \caption{Evolution of the score difference in a typical NBA game: the
    Denver Nuggets vs.\ the Chicago Bulls on 26 November 2010.  Dots indicate
    the four lead changes in the game.  The Nuggets led for 2601 out of 2880
    total seconds and won the game by a score of 98--97.}
\label{example}
\end{figure}

To help understand scoring dynamics in team sports and to set the stage for
our theoretical approach, we outline basic observations 
about NBA regular-season
games. 
In an average game, the
two teams combine to score an average of 93.6 baskets (Table~\ref{tab:data}), with an
average value of 2.07 points per basket (the point value greater than 2
arises because of foul shots and 3-point baskets).  The
average scores of the winning and losing teams are 102.1 and 91.7 points,
respectively, so that the total average score is 193.8 points in a 48-minute
game ($T\!=\!2880$ seconds).  The rms score difference between the winning
and losing teams is 13.15 points.  The high scoring rate in basketball
provides a useful laboratory to test our random-walk description of
scoring (Fig.~\ref{example}).

Scoring in professional basketball has several
additional important features~\cite{gabel:redner:2012,merritt:clauset:2014}:
\begin{enumerate}
\itemsep -0.25ex

\item Nearly constant scoring rate throughout the game, except for small
  reductions at the start of the game and the second half, and a substantial
  enhancement in the last 2.5 minutes.

\item Essentially no temporal correlations between successive scoring events.

\item Intrinsically different team strengths.  This feature may be modeled by
  a bias in the underlying random walk that describes scoring.

\item Scoring antipersistence.  Since the team that scores cedes ball
  possession, the probability that this team again scores next occurs with
  probability $p<\frac{1}{2}$.

\item Linear restoring bias.  On average, the losing team scores at a
  slightly higher rate than the winning team, with the rate disparity
  proportional to the score difference.
\end{enumerate}

A major factor for the scoring rate is the 24-second ``shot clock,'' in which
a team must either attempt a shot that hits the rim of the basket within 24
seconds of gaining ball possession or lose its possession.  The average time
interval between scoring events is $\Delta t=2880/93.6=30.8$ seconds, consistent with
the 24-second shot clock.  In a random walk picture of scoring, the average
number of scoring events in a game, $N\!=\!93.6$, together with $s\!=\!2.07$ points
for an average event, would lead to an rms displacement of
$x_{\rm rms} = \sqrt{Ns^2}$.  However, this estimate does not account for the
antipersistence of basketball scoring.  Because a team that scores
 immediately cedes ball possession, the probability that this same team
scores next occurs with probability $p\approx 0.36$.  This antipersistence
reduces the diffusion coefficient of a random walk by a factor
$p/(1-p)\approx 0.562$~\cite{gabel:redner:2012,garcia:2007}.  Using this, we
infer that the rms score difference in an average basketball game should be
$\Delta S_{\rm rms}\approx \sqrt{pNs^2/(1-p)}\approx 15.01$ points.  Given
the crudeness of this estimate, the agreement with the empirical value of
13.15 points is satisfying.

A natural question is whether this final score difference is determined by
random-walk fluctuations or by disparities in team strengths.  As we now
show, for a typical game, these two effects have comparable influence.  The
relative importance of fluctuations to systematics in a stochastic process is
quantified by the P\'eclet number $\mathcal{P}\!e \!\equiv\! vL/2D$~\cite{P94},
where $v$ is the bias velocity, $L=vT$ is a characteristic final score
difference, and $D$ is the diffusion coefficient.  Let us now estimate the
P\'eclet number for NBA basketball.  Using $\Delta S_{\rm rms}= 13.15$
points, we infer a bias velocity $v=13.15/2880\approx 0.00457$ points/sec
under the assumption that this score difference is driven only by the
differing strengths of the two competing teams.  We also estimate the
diffusion coefficient of basketball as
$D = \frac{p}{1-p} (s^2/2\Delta t)\approx 0.0391$ (points)$^2$/sec.  With
these values, the P\'eclet number of basketball is
\begin{equation}
\label{Pe}
\mathcal{P}\!e= \frac{vL}{2D}\approx 0.77 \enspace .
\end{equation}
Since the P\'eclet number is of the order of 1, systematic effects do not
predominate, which accords with common experience---a team with a weak
win/loss record on a good day can beat a team with a strong record on a bad
day.  Consequently, our presentation on scoring statistics is mostly based on
the assumption of equal-strength teams.  However, we also discuss the case of
unequal team strengths for logical completeness.

As we will present below, the statistical properties of lead changes and lead
magnitudes, and the probability that a lead is ``safe,'' i.e., will not be
erased before the game is over, are well described by an unbiased random-walk
model.  The agreement between the model predictions and data is closest for
basketball.  For the other professional sports, some discrepancies with the
random-walk model arise that may help identify alternative mechanisms for
scoring dynamics.

\section{Number of Lead Changes and Fraction of Time Leading}

Two simple characterizations of leads are:\ (i) the average number
of lead changes $\mathcal{N}$ in a game, and (ii) the fraction of game time
that a randomly selected team holds the lead.  We define a lead change as an
event where the score difference returns to zero (i.e., a tie score), but do
not count the initial score of 0--0 as lead change.  We estimate the number
of lead changes by modeling the evolution of the score difference as an
unbiased random walk.

Using $N=93.6$ scoring events per game, together with the well-known
probability that an $N$-step random walk is at the origin, the random-walk
model predicts $\sqrt{2N/\pi}\approx 8$ for a typical number of lead changes.
Because of the antipersistence of basketball scoring, the above is an
underestimate.  More properly, we must account for the reduction of the
diffusion coefficient of basketball by a factor of $p/(1-p)\approx 0.562$
compared to an uncorrelated random walk.  This change increases the number of
lead changes by a factor $1/\sqrt{0.562}\approx 1.33$, leading to roughly
10.2 lead changes.  This crude estimate is close to the observed 9.4 lead
changes in NBA games (Table~\ref{tab:data}).

\begin{figure}[ht]
  \centerline{\includegraphics[width=0.45\textwidth]{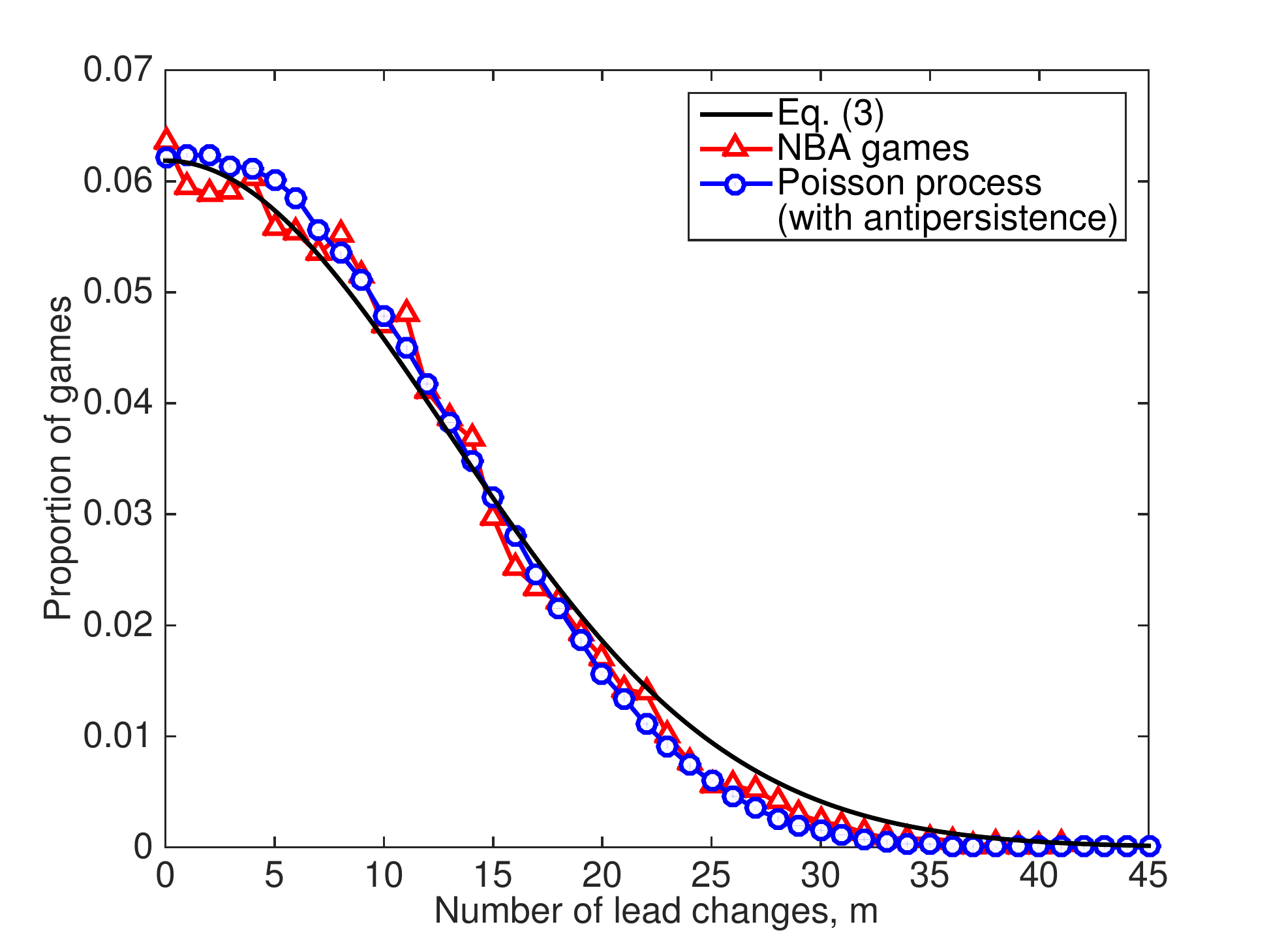}}
  \caption{Distribution of the average number of lead changes per game in
    professional basketball.}
\label{NL}
\end{figure}

For the distribution of the number of lead changes, we make use of the 
well-known result that the probability $G(m,N)$ that a discrete $N$-step random
walk makes $m$ returns to the origin asymptotically has the Gaussian form
$G(m,N)\sim e^{-m^2/2N}$~\cite{weiss:rubin:1983,weiss:1994,redner:2001}.
However, the antipersistence of basketball scoring leads to $N$ being
replaced by $N\frac{1-p}{p}$, so that the probability of making $m$ returns to
the origin is given by
\begin{equation}
\label{G}
G(m,N) \simeq \sqrt{\frac{2p}{\pi N(1-p)}}\,\, e^{-m^2p/[2N(1-p)]} \enspace .
\end{equation}
Thus $G(m,N)$ is broadened compared to the uncorrelated random-walk
prediction because lead changes now occur more frequently.  The comparison
between the empirical NBA data for $G(m,N)$ and a simulation in which scoring
events occur by an antipersistent Poisson process (with average scoring rate
of one event every 30.8 seconds), and Eq.~\eqref{G} is given in
Fig.~\ref{NL}.

For completeness, we now analyze the statistics of lead changes for unequally
matched teams.  Clearly, a bias in the underlying random walk for scoring
events decreases the number of lead changes.  We use a suitably adapted
continuum approach to estimate the number of lead changes in a simple way.
We start with the probability that biased diffusion lies in a small range
$\Delta x$ about $x=0$:
\begin{equation*}
 \frac{\Delta x}{\sqrt{4\pi Dt}} \,\, e^{-v^2t/4D} \enspace .
\end{equation*}
Thus the local time that this process spends within $\Delta x$ about the
origin up to time $t$ is
\begin{align}
\mathcal{T}(t) &=\Delta x \int_0^t \frac{dt}{\sqrt{4\pi Dt}} \,\, e^{-v^2t/4D} \nonumber \\
 &=\frac{\Delta x}{v}\,\,\mathrm{erf}\big(\sqrt{{v^2t}/{4D}}\big) \enspace ,
\end{align}
where we used $w=\sqrt{v^2t/4D}$ to transform the first line into the
standard form for the error function.  To convert this local time to number
of events, $\mathcal{N}(t)$, that the walk remains within $\Delta x$, we
divide by the typical time $\Delta t$ for a single scoring event.  Using
this, as well as the asymptotics of the error function, we obtain the
limiting behaviors:
\begin{equation}
\label{Nt}
\mathcal{N}(t)=\frac{\mathcal{T}(t)}{\Delta t} \sim 
\begin{cases} \sqrt{{v_0^2t}/{\pi D}} & \qquad {v^2t}/{4D}\ll 1\\
  {v_0}/{v}  & \qquad {v^2t}/{4D}\gg 1 \enspace ,
\end{cases}
\end{equation}
with $v_0={\Delta x}/{\Delta t}$ and $\Delta x$ the average value of a single
score (2.07 points).  Notice that $v^2T/4D=\mathcal{P}\!e/2$, which, from
Eq.~\eqref{Pe}, is roughly 0.38.  Thus, for the NBA, the first line of
Eq.~\eqref{Nt} is the realistic case.  This accords with what we have already
seen in Fig.~\ref{NL}, where the distribution in the number of lead changes
is accurately accounted for by an unbiased, but antipersistent random walk.

\begin{figure}[t!]
\centerline{\includegraphics[width=0.45\textwidth]{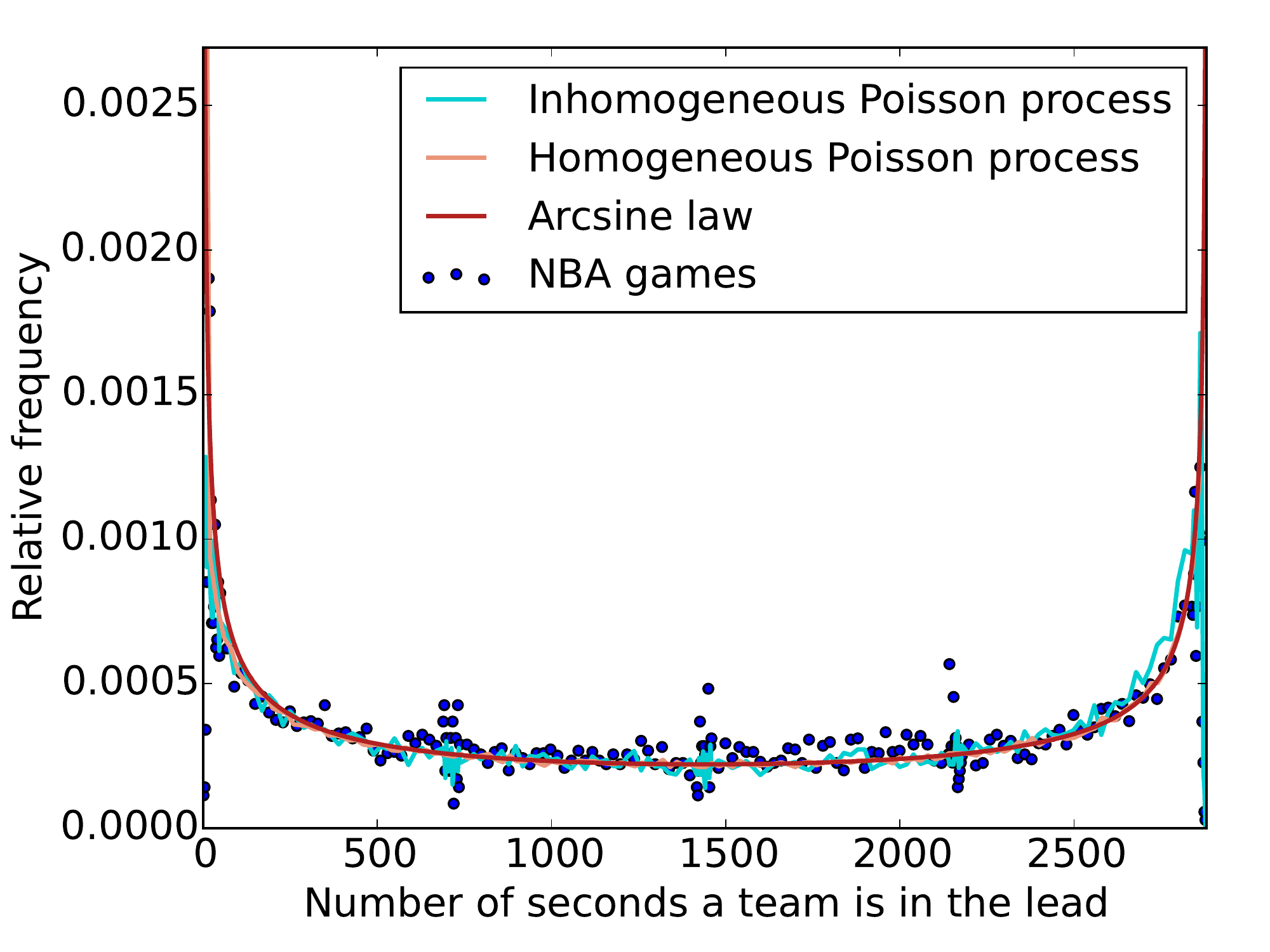}}
\caption{The distribution of the time that a given team holds the lead,
  $\mathcal{O}(t)$. }
\label{fig:first:arcsin}
\end{figure}

Another basic characteristic of lead changes is the amount of game time that
one team spends in the lead, $\mathcal{O}(t)$, a quantity that has been
previously studied for basketball~\cite{gabel:redner:2012}.  Strikingly, the
probability distribution for this quantity is bimodal, in which
$\mathcal{O}(t)$ sharply increases as the time approaches either 0 or $T$,
and has a minimum when the time is close to ${T}/{2}$.  If the scoring
dynamics is described by an unbiased random walk, then the probability that
one team leads for a time $t$ in a game of length $T$ is given by the first
arcsine law of Eq.~\eqref{arcsine}~\cite{feller:1968,redner:2001}.
Figure~\ref{fig:first:arcsin} compares this theoretical result with
basketball data.  Also shown are two types of synthetically generated data.
For the ``homogeneous Poisson process'', we use the game-averaged scoring
rate to generate synthetic basketball-game time series of scoring events.
For the ``inhomogeneous Poisson process'', we use the empirical instantaneous scoring
rate for each second of the game to generate the synthetic data
(Fig.~\ref{Lt}).  As we will justify in the next section, we do not
incorporate the antipersistence of basketball scoring in these Poisson
processes because this additional feature minimally influences the
distributions that follow the arcsine law ($\mathcal{O}$, $\mathcal{L}$ and
$\mathcal{M}$).  The empirically observed increased scoring rate at the end
of each quarter~\cite{gabel:redner:2012,merritt:clauset:2014}, leads to
anomalies in the data for $\mathcal{O}(t)$ that are accurately captured by
the inhomogeneous Poisson process.

\section{Time of the Last Lead Change}

We now determine \emph{when} the last lead change occurs.  For the discrete
random walk, the probability that the last lead occurs after $N$ steps can be
solved by exploiting the reflection principle~\cite{feller:1968}.  Here we
solve for the corresponding distribution in continuum diffusion because this
formulation is simpler and we can readily generalize to unequal-strength
teams.  While the distribution of times for the last lead change is well
known~\cite{levy:1939,levy:1948}, our derivation is intuitive and elementary.

\begin{figure}[h!]
\centerline{\includegraphics[width=0.325\textwidth]{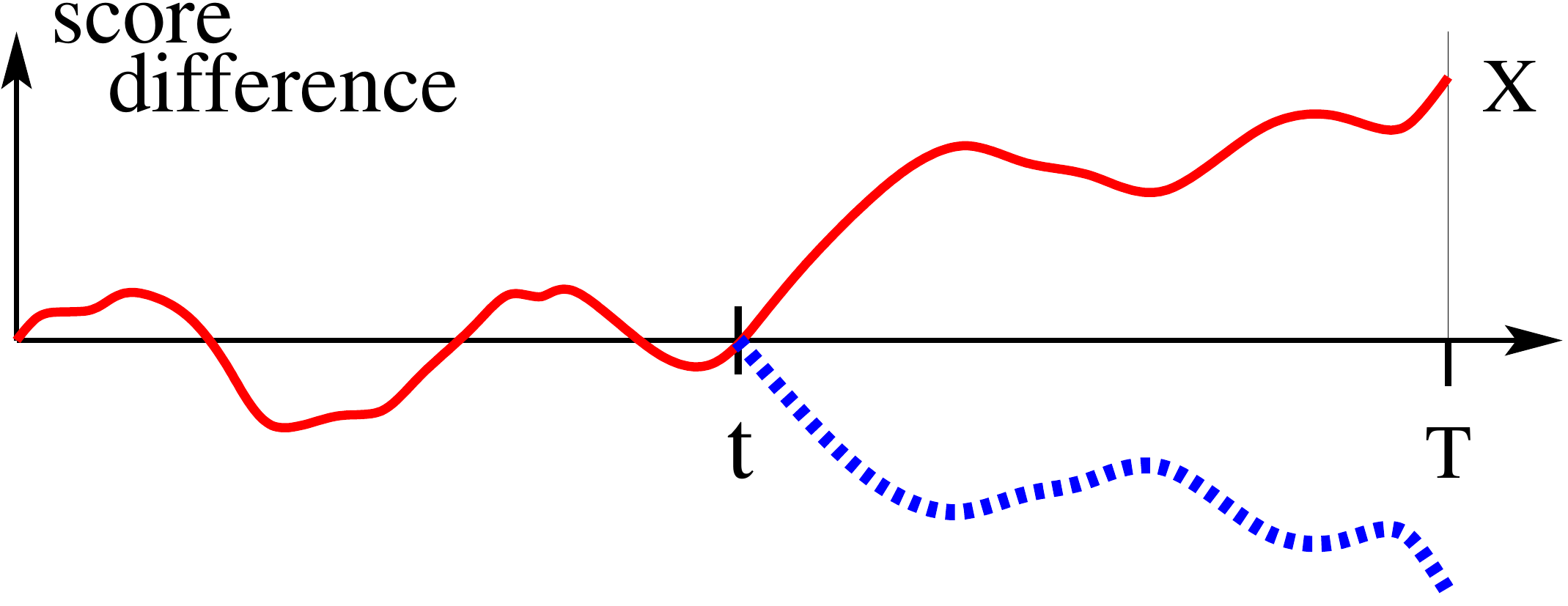}}
\caption{Schematic score evolution in a game of time $T$.  The subsequent
  trajectory after the last lead change must always be positive (solid) or
  always negative (dashed).}
\label{last-time}
\end{figure}

For the last lead change to occur at time $t$, the score difference, which
started at zero at $t\!=\!0$, must again equal zero at time $t$
(Fig.~\ref{last-time}).  For equal-strength teams, the probability for this
event is simply the Gaussian probability distribution of diffusion evaluated
at $x\!=\!0$:
\begin{equation}
\label{P0}
P(0,t)= \frac{1}{\sqrt{4\pi Dt}} \enspace .
\end{equation}
To guarantee that it is the \emph{last} lead change that occurs at time $t$,
the subsequent evolution of the score difference, cannot cross the origin
between times $t$ and $T$ (Fig.~\ref{last-time}).  To enforce this
constraint, the remaining trajectory between $t$ and $T$ must therefore be a
\emph{time-reversed} first-passage path from an arbitrary final point $(X,T)$
to $(0,t)$.  The probability for this event is the first-passage
probability~\cite{redner:2001}
\begin{equation}
\label{fpp}
F(X,T\!-\!t)= \frac{X}{\sqrt{4\pi D(T\!-\!t)^3}}\,\,e^{-X^2/[4D(T-t)]} \enspace .
\end{equation}

\begin{figure}[t!]
\centerline{\includegraphics[width=0.45\textwidth]{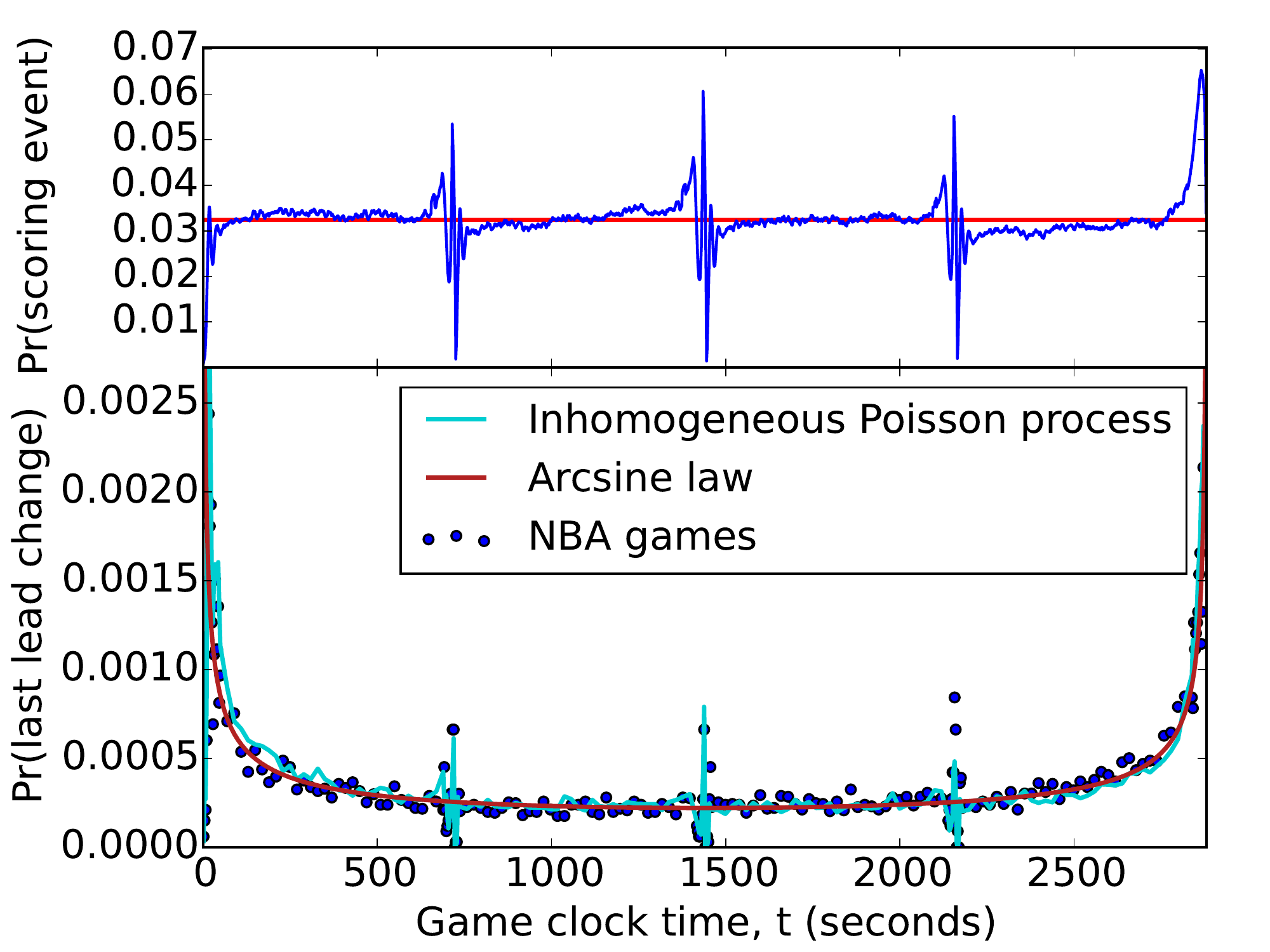}}
\caption{(upper) Empirical probability that a scoring event occurs at time
  $t$, with the game-average scoring rate shown as a horizontal line.   The
  data is aggregated in bins of 10 seconds each; the same binning is used in Fig.~\ref{fig:last:lead:change}. (lower)
Distribution of times $\mathcal{L}(t)$ for the last lead change.}
\label{Lt}
\end{figure}

With these two factors, the probability that the last lead change occurs at
time $t$ is given by
\begin{align}
\label{Pi}
\mathcal{L}(t)&=2\!\int_0^\infty \!dX\, P(0,t)\, F(X,T\!-\!t) \nonumber \\
& = 2\!\int_0^\infty \!\!\frac{dX}{\sqrt{4\pi Dt}} \frac{X}{\sqrt{4\pi D(T\!-\!t)^3}}\,\,
e^{-X^2/[4D(T\!-\!t)]} \,.
\end{align}
The leading factor 2 appears because the subsequent trajectory after time $t$
can equally likely be always positive or always negative.  The integration is
elementary and the result is the classic second arcsine
law~\cite{levy:1939,levy:1948} given in Eq.~\eqref{arcsine}.  The salient
feature of this distribution is that the last lead change in a game between
evenly matched teams is most likely to occur either near the start or the end
of a game, while a lead change in the middle of a game is less likely. 

As done previously for the distribution of time $\mathcal{O}(t)$ that one
team is leading, we again generate a synthetic time series that is based on a
homogeneous and an inhomogeneous Poisson process for individual scoring
events without antipersistence.  From these synthetic histories, we extract
the time for the last lead and its distribution.  The synthetic inhomogeneous
Poisson process data accounts for the end-of-quarter anomalies in the
empirical data with remarkable accuracy (Fig.~\ref{Lt}).

\begin{figure}[t!]
  \centerline{\includegraphics[width=0.45\textwidth]{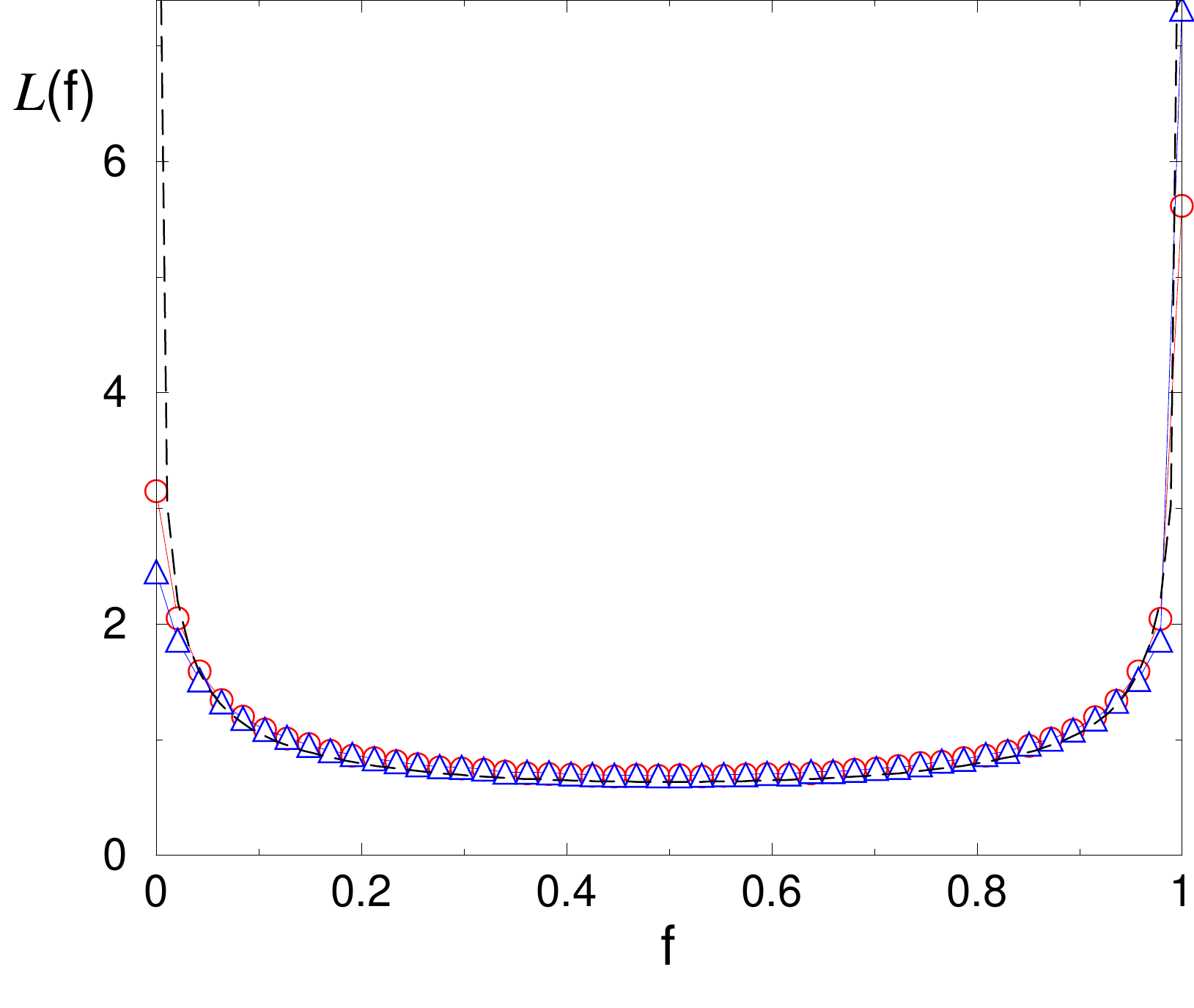}}
  \caption{The distribution of time for the last lead change,
    $\mathcal{L}(f)$, as a function of the fraction of steps $f$ for a
    94-step random walk with persistence parameter $p\!=\!0.36$ as in the NBA
    ($\circ$) and $p\!=\!0.25$ corresponding to stronger persistence
    ($\bigtriangleup$).  The smooth curve is the arcsine law for $p\!=\!0.5$
    (no antipersistence). }
\label{LLA}
\end{figure}

Let us now investigate the role of scoring antipersistence on the
distribution $\mathcal{L}(t)$.  While the antipersistence substantially
affects the number of lead changes and its distribution, antipersistence has
a barely perceptible effect on $\mathcal{L}(t)$.  Figure~\ref{LLA} shows the
probability $\mathcal{L}(f)$ that the last lead change occurs when a fraction
$f$ of the steps in an $N$-step antipersistent random walk have occurred,
with $N\!=\!94$, the closest even integer to the observed value $N\!=\!93.56$
of NBA basketball.  For the empirical persistence parameter of basketball,
$p=0.36$, there is little difference between $\mathcal{L}(f)$ as given by the
arcsine law and that of the data, except at the first two and last two steps
of the walk.  Similar behavior arises for the more extreme case of
persistence parameter $p=0.25$.  Thus basketball scoring antipersistence
plays little role in determining the time at which the last lead change
occurs.

We may also determine the role of a constant bias on $\mathcal{L}(t)$,
following the same approach as that used for unbiased diffusion.  Now the
analogues of Eqs.~\eqref{P0} and~\eqref{fpp} are~\cite{redner:2001},
\begin{align}
\begin{split}
\label{P0-bias}
P(0,v,t)&= \frac{1}{\sqrt{4\pi Dt}}\,\, e^{-v^2t/4D} \\
F(X,v,t)&= \frac{X}{\sqrt{4\pi Dt^3}}\,\, e^{-(X+vt)^2/4Dt}\enspace .
\end{split}
\end{align}
Similarly, the analogue of Eq.~\eqref{Pi} is
\begin{align}
\label{Pi-bias}
\mathcal{L}(t)=\int_0^\infty \!\!\!dX\, P(0,t)\,\big[ F(X,v,T\!-\!t)\!+\! F(X,-v,T\!-\!t)\big] \,.
\end{align}
In Eq.~\eqref{P0-bias} we must separately consider the situations where the
trajectory for times beyond $t$ is strictly positive (stronger team
ultimately wins) or strictly negative (weaker team wins).  In the former
case, the time-reversed first-passage path from $(X,T)$ to $(0,t)$ is
accomplished in the presence of a positive bias $+v$, while in the latter
case, this time-reversed first passage occurs in the presence of a negative
bias $-v$.

\begin{figure}[t!]
 \centerline{\includegraphics[width=0.45\textwidth]{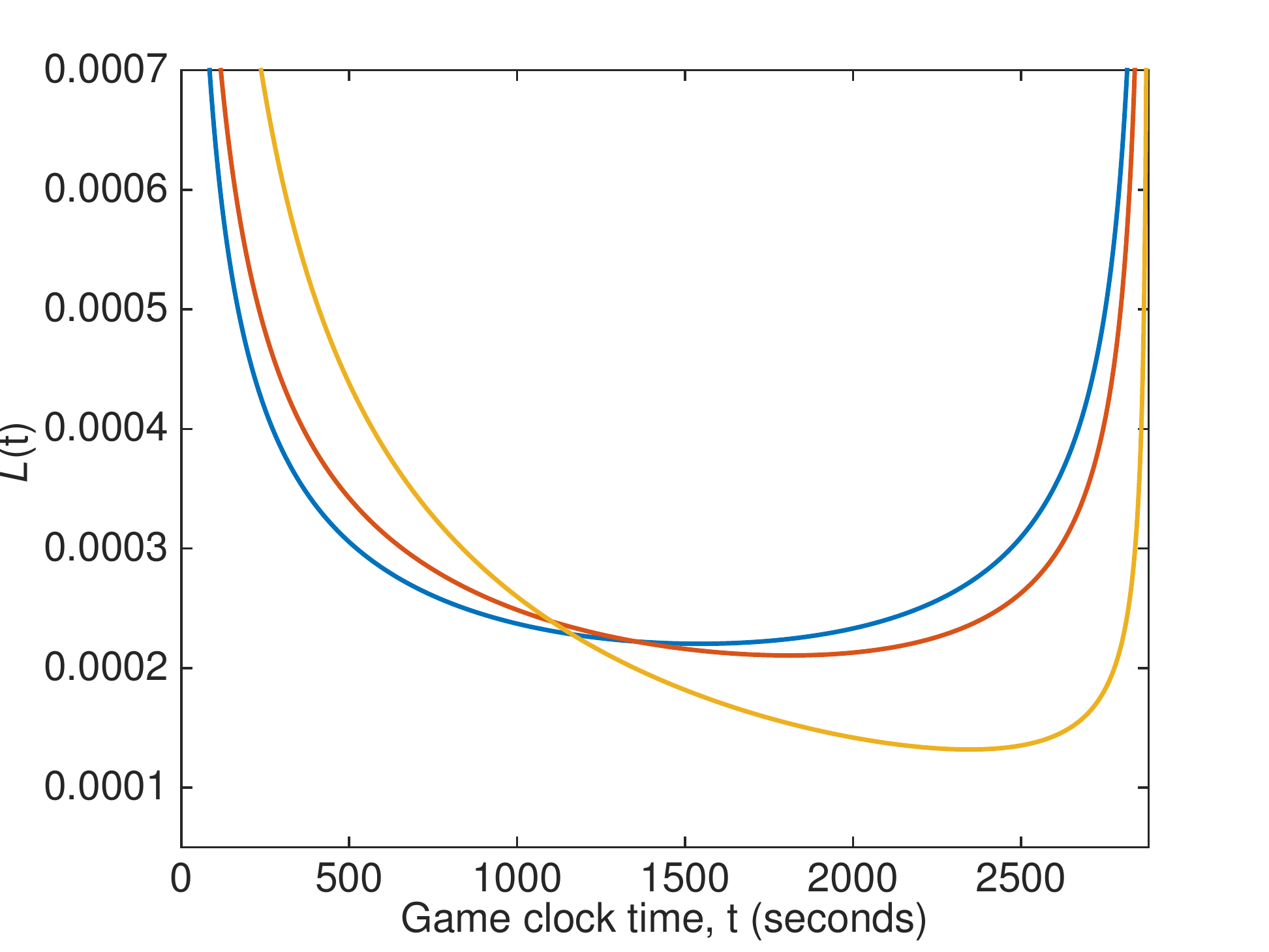}}
 \caption{The distribution $\mathcal{L}(t)$ for non-zero bias
   (Eq.~\eqref{Lt-v}).  The diffusion coefficient is the empirical value
   $D=0.0391$, and bias values are: $v=0.002$, $v=0.004$, and $v=0.008$
   (increasingly asymmetric curves).  The central value of $v$ roughly
   corresponds to average NBA game-scoring bias if diffusion is neglected.}
\label{Lv}
\end{figure}

Explicitly, Eq.~\eqref{Pi-bias} is
\begin{align}
\mathcal{L}(t)&=  \frac{e^{-v^2t/4D}}{\sqrt{4\pi Dt}}
\int_0^\infty \!\!\!dX \frac{X}{\sqrt{4\pi D(T\!-\!t)^3}}\,\,\nonumber \\
&\hskip 0.7in \times \left\{e^{-(X+a)^2/b} + e^{-(X-a)^2/b}\right\}\,,
\end{align}
with $a=v(T-t)$ and $b=4D(T-t)$.  Straightforward calculation gives
\begin{align}
\label{Lt-v}
\mathcal{L}(t)
&= \frac{e^{-v^2t/4D}}{\pi\sqrt{t(T\!-\!t)}} 
\left\{\sqrt{\frac{\pi v^2(T\!-\!t)}{4D}} \,\, 
\mathrm{erf}\bigg(\sqrt{\frac{v^2(T\!-\!t)}{4D}}\,\bigg)\right.\nonumber \\
&\hskip 1.3in \left. +e^{-v^2(T\!-\!t)/4D}\right\}\,.
\end{align}
This form for $\mathcal{L}(t)$ is again bimodal (Fig.~\ref{Lv}), as in the
arcsine law, but the last lead change is now more likely to occur near the
beginning of the game.  This asymmetry arises because once a lead is
established, which is probable because of the bias, the weaker team is
unlikely to achieve another tie score.

More germane to basketball, we should average $\mathcal{L}(t)$ over the
distribution of biases in all NBA games.  For this averaging, we use the
observation that many statistical features of basketball are accurately
captured by employing a Gaussian distribution of team strengths with mean
value 1 (since the absolute strength is immaterial), and standard deviation
of approximately 0.09~~\cite{gabel:redner:2012}.  This parameter value was
inferred by using the Bradley-Terry competition
model~\cite{bradley:terry:1952}, in which teams of strengths $S_1$ and $S_2$
have scoring rates $S_1/(S_1+S_2)$ and $S_2/(S_1+S_2)$, respectively, to
generate synthetic basketball scoring time series.  The standard deviation
0.09 provided the best match between statistical properties that were
computed from the synthetic time series and the empirical game
data~\cite{gabel:redner:2012}.  From the distribution of team strengths, we
then infer a distribution of biases for each game and finally average over
this bias distribution to obtain the bias-averaged form of $\mathcal{L}(t)$.
The skewness of the resulting distribution is minor and it closely matches the bias-free
form of $\mathcal{L}(t)$ given in Fig.~\ref{Lt}.  Thus, the bias of
individual games appears to again play a negligible role in statistical properties
of scoring, such as the distribution of times for the last lead change.

\section{Time of the Maximal Lead}

We now ask when the \emph{maximal} lead occurs in a
game~\cite{majumdar:etal:2008}.  If the score difference evolves by unbiased
diffusion, then the standard deviation of the score difference grows as
$\sqrt{t}$.  Naively, this behavior might suggest that the maximal lead
occurs near the end of a game.  In fact, however, the probability
$\mathcal{M}(t)$ that the maximal lead occurs at time $t$ also obeys the
arcsine law Eq.~\eqref{arcsine}.  Moreover, the arcsine laws for the last
lead time and for the maximal lead time are
equivalent~\cite{levy:1939,levy:1948,morters:peres:2010}, so that the largest
lead in a game between two equally-matched teams is most likely to occur
either near the start or near the end of a game.

\begin{figure}[ht]
  \centerline{\includegraphics[width=0.35\textwidth]{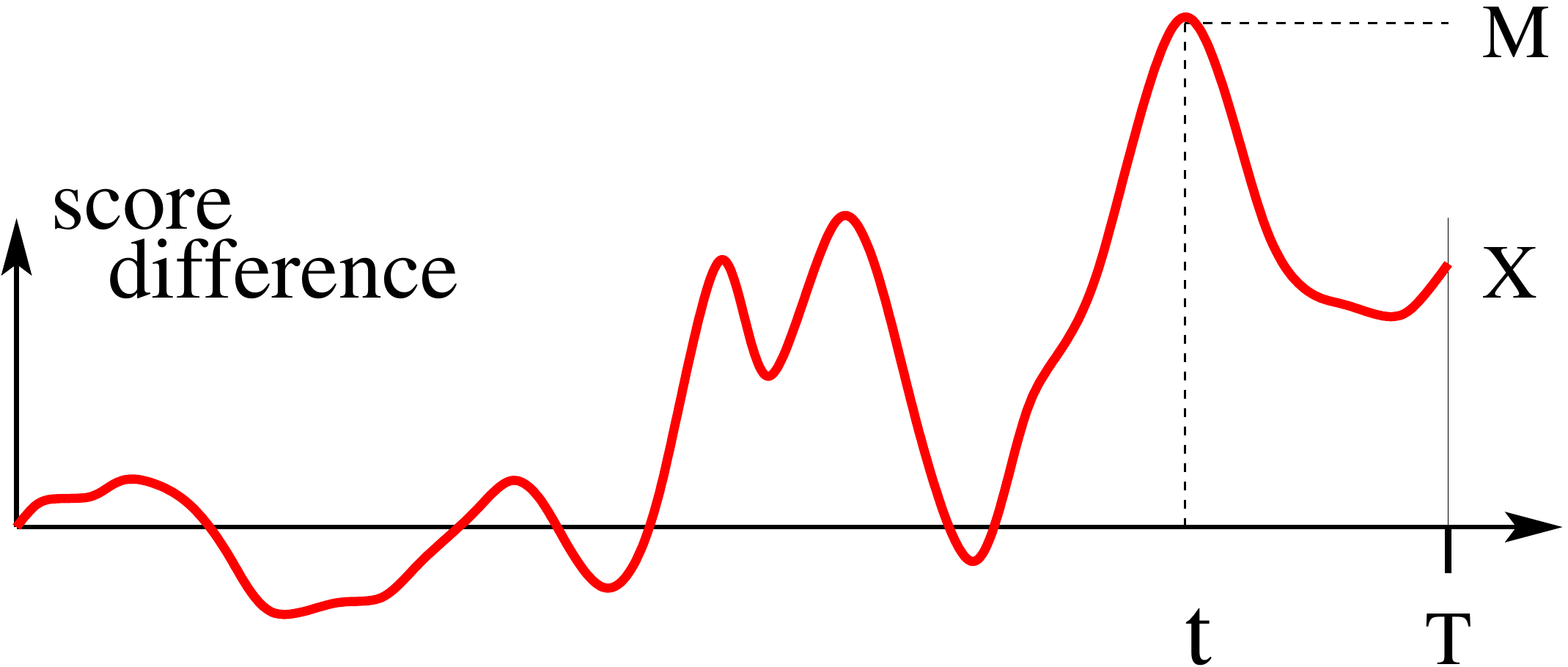}}
  \caption{ The maximal lead (which could be positive or negative) occurs at
    time $t$.}
\label{max-time}
\end{figure}

For completeness, we sketch a derivation for the distribution
$\mathcal{M}(t)$ by following the same approach used to find
$\mathcal{L}(t)$.  Referring to Fig.~\ref{max-time}, suppose that the maximal
lead $M$ occurs at time $t$.  For $M$ to be a maximum, the initial trajectory
from $(0,0)$ to $(M,t)$ must be a first-passage path, so that $M$ is never
exceeded prior to time $t$.  Similarly, the trajectory from $(M,t)$ to the
final state $(X,T)$ must also be a time-reversed first-passage path from
$(X,T)$ to $(M,t)$, but with $X<M$, so that $M$ is never exceeded for all
times between $t$ and $T$.

\begin{figure}[ht]
\centerline{\includegraphics[width=0.45\textwidth]{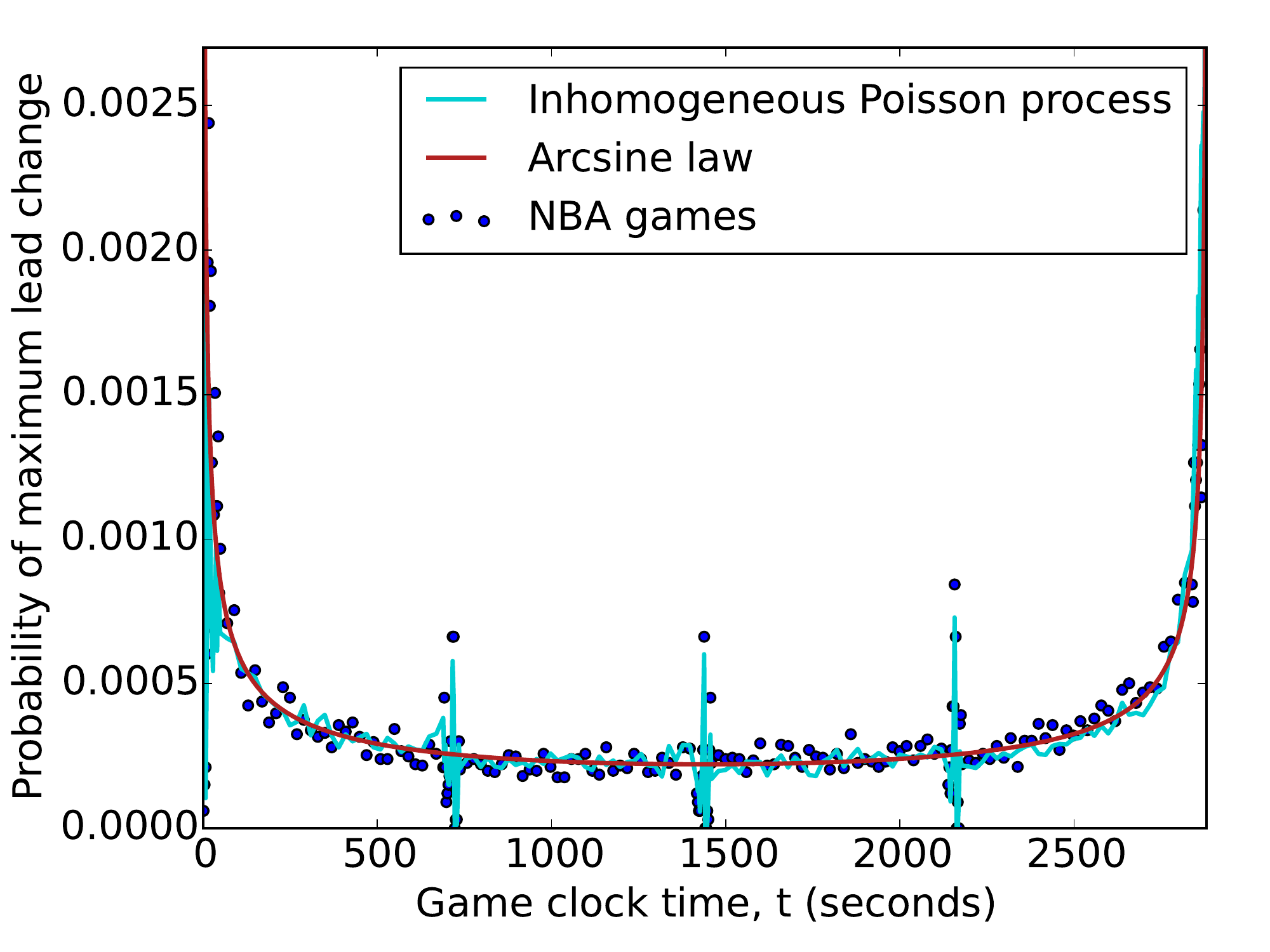}}
\caption{Distribution of times $\mathcal{M}(t)$ for the maximal lead.  }
\label{Mt}
\end{figure}

Based on this picture, we may write $\mathcal{M}(t)$ as
\begin{align}
\label{eq:Mt}
  \mathcal{M}(t)&=A
        \int_0^\infty \! dM\, F(M,t)\int_{-\infty}^M \! dX\, F(X\!-\!M,T\!-\!t)\nonumber \\
   &=A \int_0^\infty \! dM \frac{M}{\sqrt{4\pi Dt^3}}\,\,   e^{-M^2/4Dt}\nonumber \\
&\hskip 0.25in \times  
\int_{-\infty}^M \! dX  \frac{(M\!-\!X)}{\sqrt{4\pi D(T\!-\!t)^3}}\,\, e^{-(M\!-\!X)^2/4D(T\!-\!t)} \,.
\end{align}
The constant $A$ is determined by the normalization condition
$\int_0^T \mathcal{M}(t)dt \!=\!1$.  Performing the above two elementary
integrations yields again the arcsine law of Eq.~\eqref{arcsine}.
Figure~\ref{Mt} compares the arcsine law prediction with empirical data from
the NBA.

\section{Probability that a Lead is Safe}

\begin{figure}[ht]
\centerline{\includegraphics[width=0.35\textwidth]{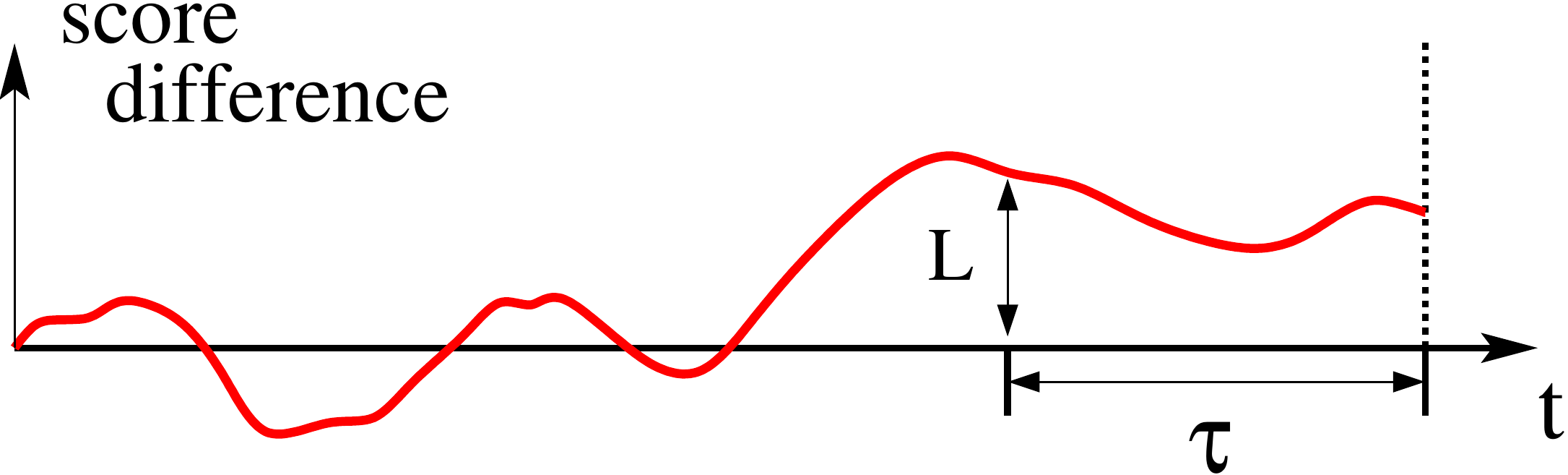}}
\caption{One team leads by $L$ points when a time $\tau$ is left in the
  game.}
\label{score-diff}
\end{figure}

Finally, we turn to the question of how safe is a lead of a given size at any
point in a game (Fig.~\ref{score-diff}), i.e., the probability that the team
leading at time $t$ will ultimately win the game.  The probability that a
lead of size $L$ is safe when a time $\tau$ remains in the game is, in
general,
\begin{align}
\label{Q}
Q(L,\tau)&=1-\int_0^\tau \!\!F(L,t)\, dt\,.
\end{align}
where $F(L,t)$ again is the first-passage probability [Eqs.~\eqref{fpp} and
\eqref{P0-bias}] for a diffusing particle, which starts at $L$, to first
reach the origin at time $t$.  Thus the right-hand side is the probability
that the lead has not disappeared up to time $\tau$.  

First consider evenly-matched teams, i.e., bias velocity $v=0$.  We
substitute $u=L/\sqrt{4Dt}$ in Eq.~\eqref{Q} to obtain
\begin{equation}
\label{Q-final}
Q(L,\tau)=1-\frac{2}{\sqrt{\pi}}\int_z^\infty e^{-u^2}\, du 
= \mathrm{erf}(z)\enspace .
\end{equation}
Here $z\equiv L/\sqrt{4D\tau}$ is the dimensionless lead size.  When
$z\!\ll\! 1$, either the lead is sufficiently small or sufficient game time
remains that a lead of scaled magnitude $z$ is likely to be erased before the
game ends.  The opposite limit of $z\!\gg \!1$ corresponds to either a
sufficiently large lead or so little time remaining that this lead likely
persists until the end of the game. We illustrate Eq.~\eqref{Q-final} with a
simple numerical example from basketball.  From this equation, a lead of
scaled size $z\approx 1.163$ is 90\% safe.  Thus a lead of 10 points is 90\%
safe when 7.87 minutes remain in a game, while an 18-point lead at the end of
the first half is also 90\% safe~\footnote{For convenience, we note that all 90\%-safe leads in professional basketball are solutions to $L=0.4602\sqrt{\tau}$, for $\tau$ seconds remaining.}.

\begin{figure}[ht]
\includegraphics[scale=0.40]{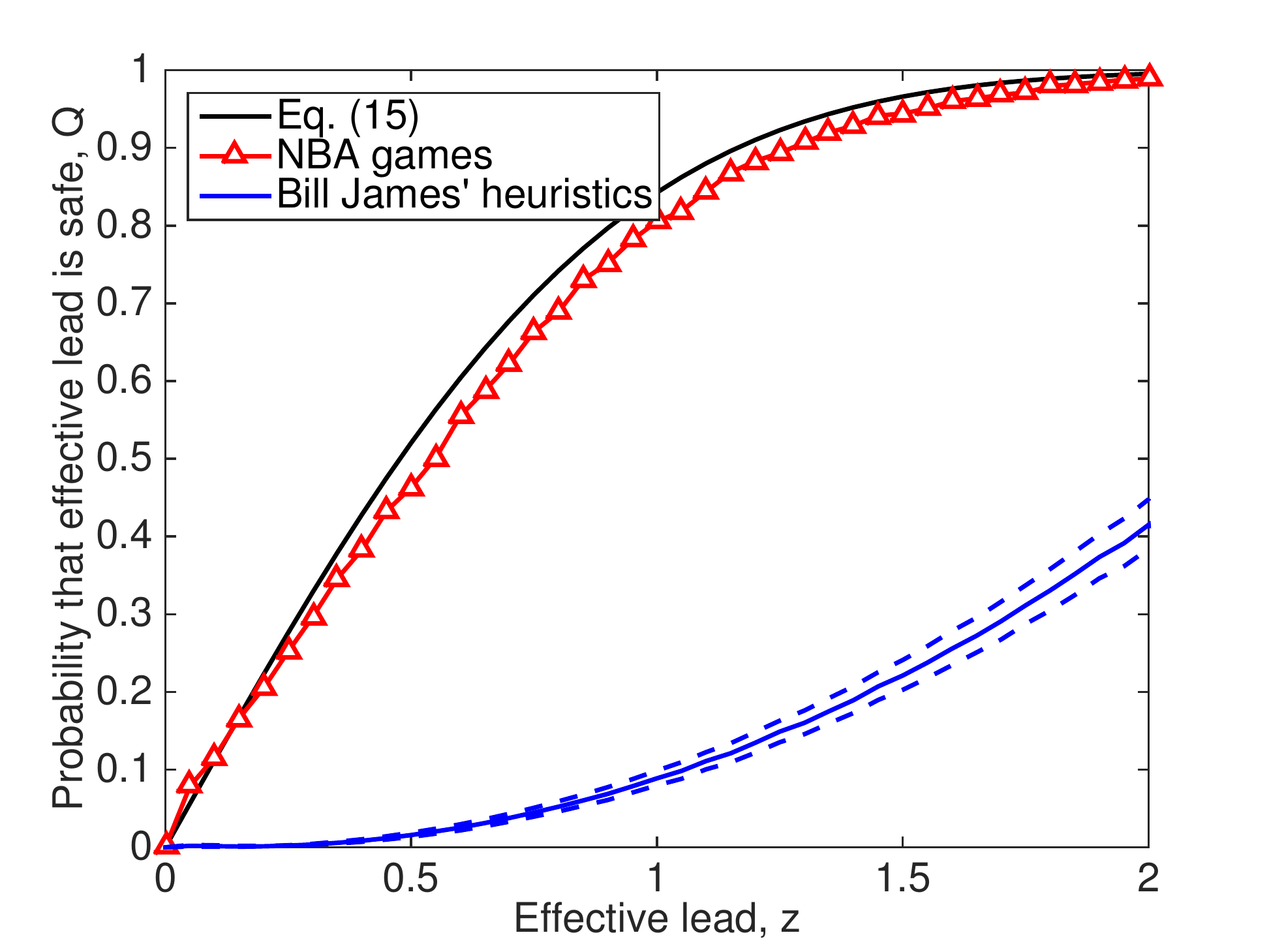}
\caption{Probability that a lead is safe versus the dimensionless lead size
  $z=L/\sqrt{4D\tau}$ for NBA games, showing the prediction from
  Eq.~\eqref{Q-final}, the empirical data, and the mean prediction for Bill
  James' well-known ``safe lead'' heuristic.}
\label{nba:safe}
\end{figure}

Figure~\ref{nba:safe} compares the prediction of Eq.~\eqref{Q-final} and the
empirical basketball data.  We also show the prediction of the heuristic
developed by basketball analyst and historian Bill James~\cite{james:2008}.
This rule is mathematically given by:\
$Q(L,\tau) = \min\left\{1,\frac{1}{\tau}(L\!-\!3\!+\!\delta/2)^2\right\}$,
where $\delta=+1$ if the leading team has ball possession and $\delta=-1$
otherwise.  The figure shows the predicted probability for
$\delta=\{-1,0,+1\}$ (solid curve for central value, dashed otherwise)
applied to all of the empirically observed $(L,\tau)$ pairs, because ball
possession is not recorded in our data.  Compared to the random walk model,
the heuristic is quite conservative (assigning large safe lead
probabilities only for dimensionless leads $z>2$) and has the wrong
qualitative dependence on $z$.  In contrast, the random walk model gives a
maximal overestimate of 6.2\% for the safe lead probability over all $z$, and
has the same qualitative $z$ dependence as the empirical data.

For completeness, we extend the derivation for the safe lead probability to
unequal-strength teams by including the effect of a bias velocity $v$ in
Eq.~\eqref{Q}:
\begin{align}
Q(L,\tau)&=1-\int_0^\tau\!\!\frac{L}{\sqrt{4\pi Dt^3}}\,\, e^{-(L+vt)^2/4Dt}\,dt \nonumber \\
&=1-e^{-vL/2D} \!\int_0^\tau\!\!\frac{L}{\sqrt{4\pi Dt^3}}\,\, e^{-L^2/4Dt-v^2t/4D}\, dt\,,
\end{align}
where the integrand in the first line is the first-passage probability for
non-zero bias.  Substituting $u=L/\sqrt{4Dt}$ and using again the P\'eclet
number $\mathcal{P}\!e= vL/2D$, the result is
\begin{align}
Q(L,\tau)&=1-\frac{2}{\sqrt{\pi}}\,\,e^{-\mathcal{P}\!e}\int_0^z
e^{-u^2-\mathcal{P}\!e^2/4u^2}\, du \nonumber\\
&=1-\tfrac{1}{2}\!\left[e^{-2\mathcal{P}\!e}\mathrm{erfc}\Big(z\!-\!\frac{\mathcal{P}\!e}{2z}\Big)
\!+\!\mathrm{erfc}\Big(z\!+\!\frac{\mathcal{P}\!e}{2z}\Big)\right] \enspace .
\end{align}
%
\begin{figure}[ht]
\subfigure[]{\includegraphics[width=0.42\textwidth]{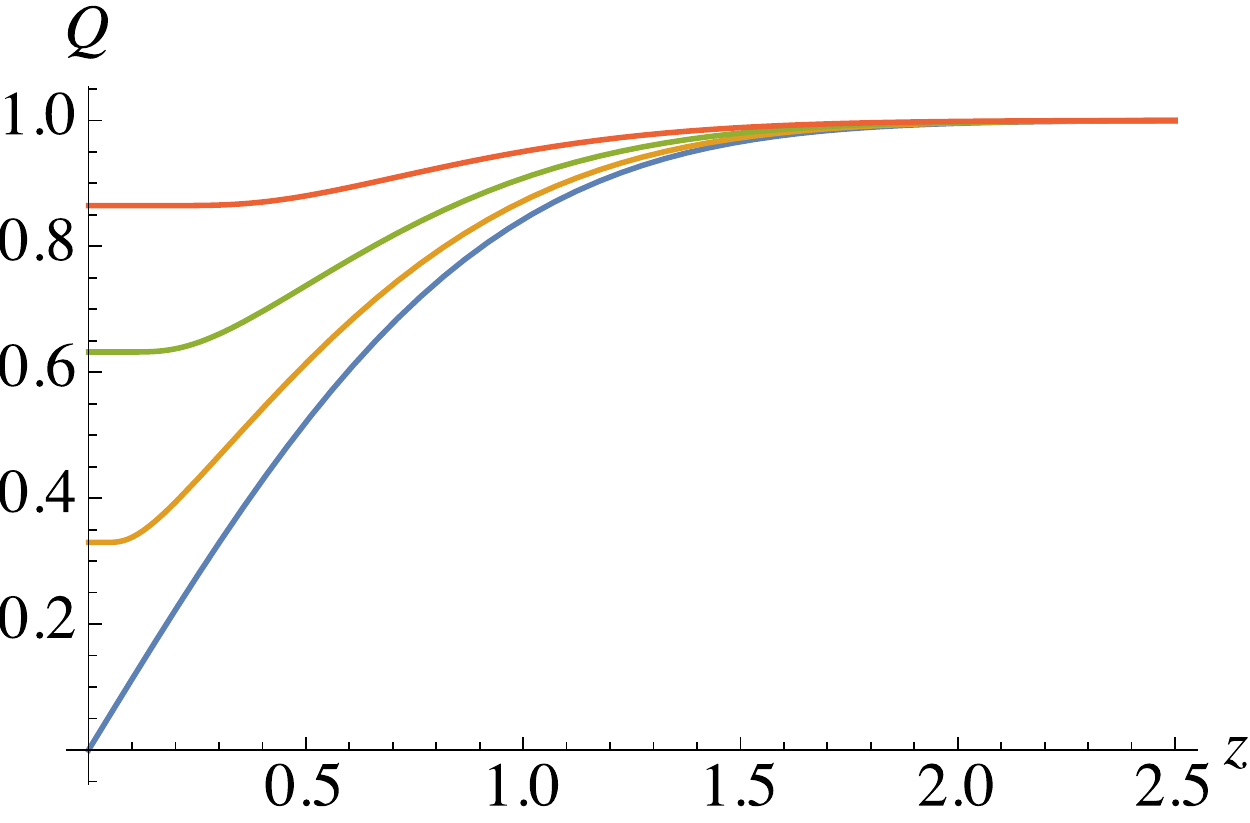}}\qquad\subfigure[]{\includegraphics[width=0.42\textwidth]{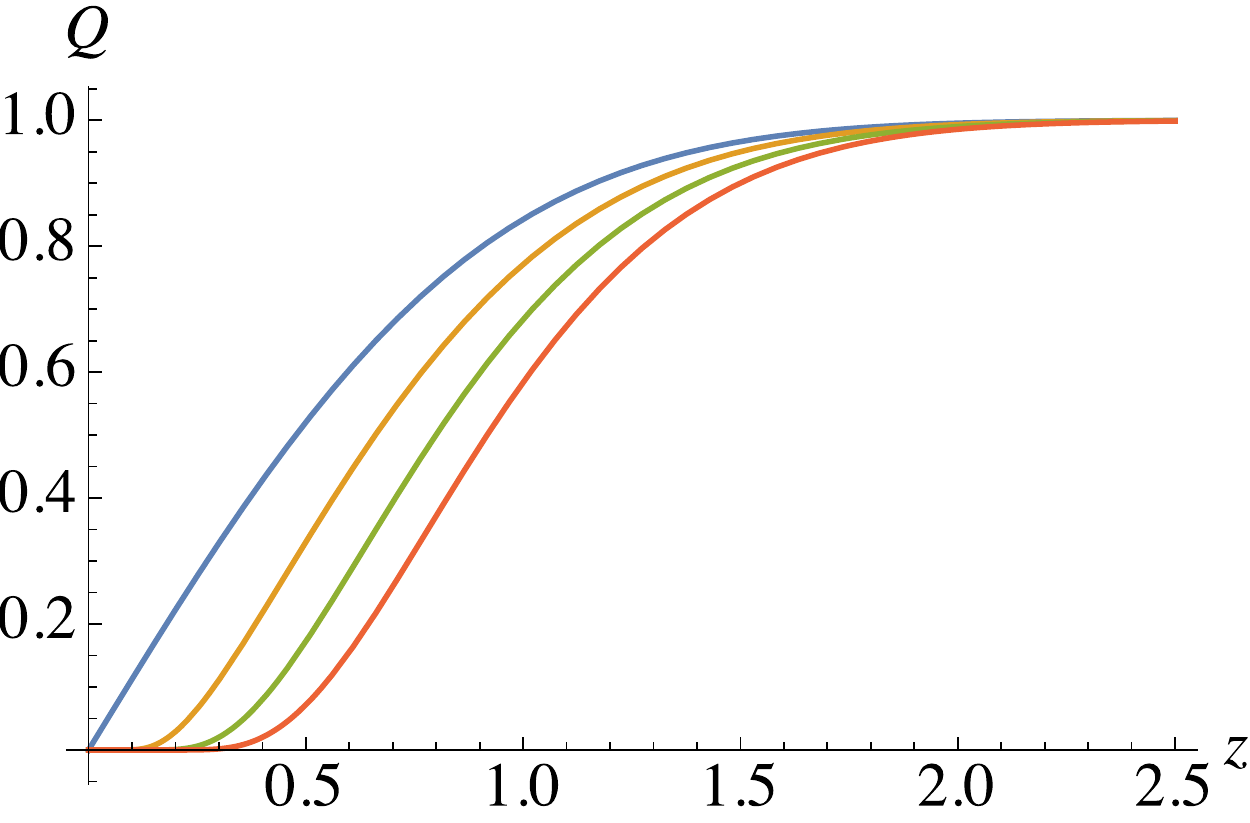}}
\caption{Probability that a lead is safe versus $z=L/\sqrt{4D\tau}$ for: (a)
  the stronger team is leading for $\mathcal{P}\!e=\frac{1}{5}$, $\frac{1}{2}$
  and 1, and (b) the weaker team is leading for $\mathcal{P}\!e=-\frac{2}{5}$,
  $-\frac{4}{5}$ and $-\frac{6}{5}$.  The case $\mathcal{P}\!e=0$ is also shown
  for comparison.}
\label{bias}
\end{figure}

When the stronger team is leading ($\mathcal{P}\!e>0$), essentially any lead is
safe for $\mathcal{P}\!e\agt 1$, while for $\mathcal{P}\!e<1$, the safety of a lead
depends more sensitively on $z$ (Fig.~\ref{bias}(a)).  Conversely, if the
weaker team happens to be leading ($\mathcal{P}\!e<0$), then the lead has to be
substantial or the time remaining quite short for the lead to be safe
(Fig.~\ref{bias}(b)).  In this regime, the asymptotics of the error function
gives $Q(L,\tau)\sim e^{-\mathcal{P}\!e^2/4z^2}$ for
$z<|\mathcal{P}\!e|/2$, which is vanishingly small.  For values of $z$
in this range, the lead is essentially never safe.

\section{Lead changes in other sports}

We now consider whether our predictions for lead change statistics in
basketball extend to other sports, such as college American football (CFB),
professional American football (NFL), and professional hockey
(NHL)~\footnote{Although the NHL data span 10 years, they include only 9
  seasons because the 2004 season was canceled as a result of a labor
  dispute.}.  These sports have the following commonalities with
basketball~\cite{merritt:clauset:2014}:
\begin{enumerate}
\itemsep -0.25ex
\item Two teams compete for a fixed time $T$, in which points are scored by
  moving a ball or puck into a special zone in the field.
\item Each team accumulates points during the game and the team with the
  largest final score is the winner (with sport-specific tiebreaking rules).
\item A roughly constant scoring rate throughout the game, except for small
  deviations at the start and end of each scoring period.
\item Negligible temporal correlations between successive scoring events.
\item Intrinsically different team strengths.
\item Scoring antipersistence, except for hockey.
\end{enumerate}
These similarities suggest that a random-walk model should also apply to lead
change dynamics in these sports.

\begin{figure*}[t!]
\begin{tabular}{ccc}
\includegraphics[scale=0.284]{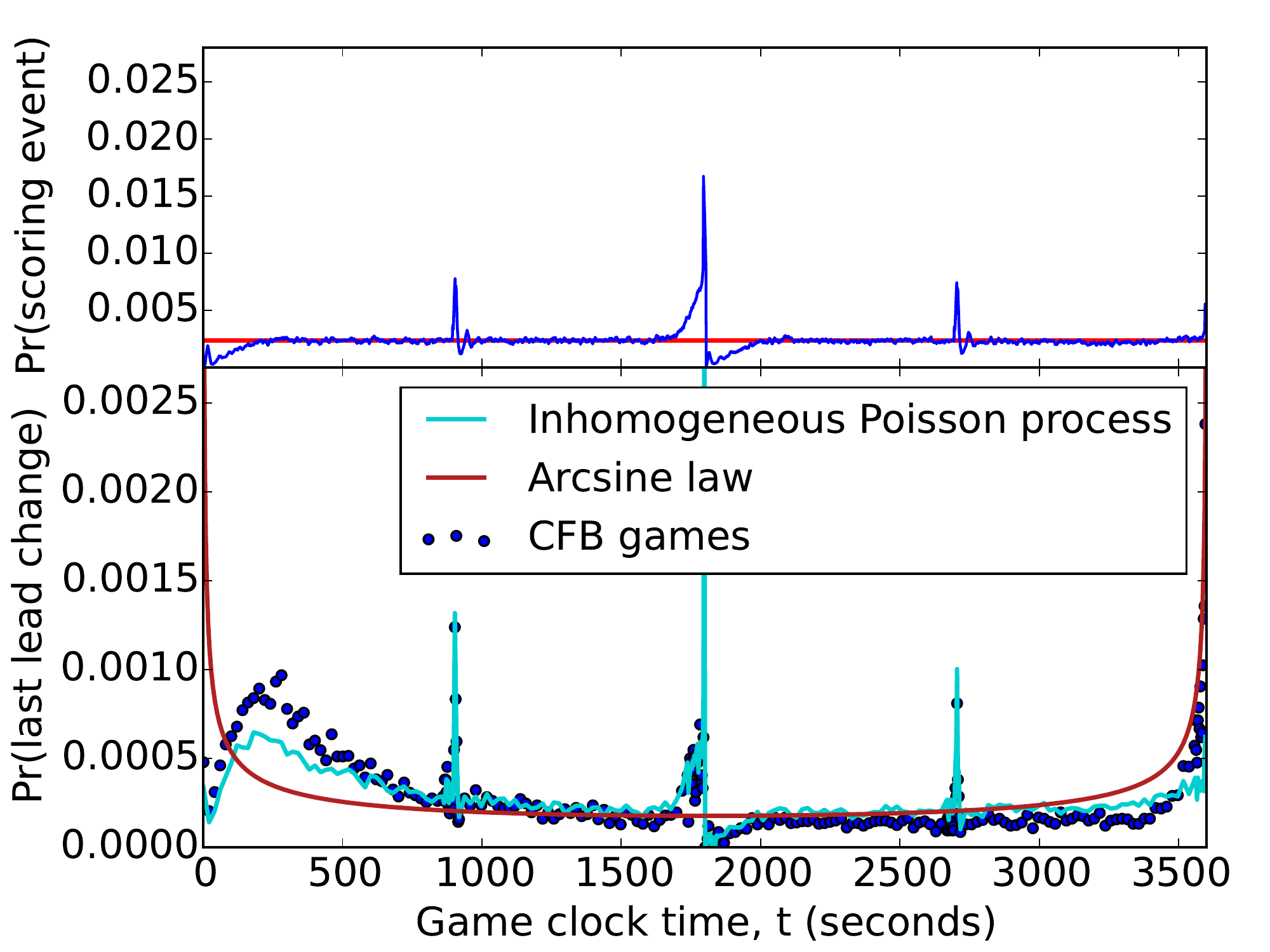} &
\includegraphics[scale=0.284]{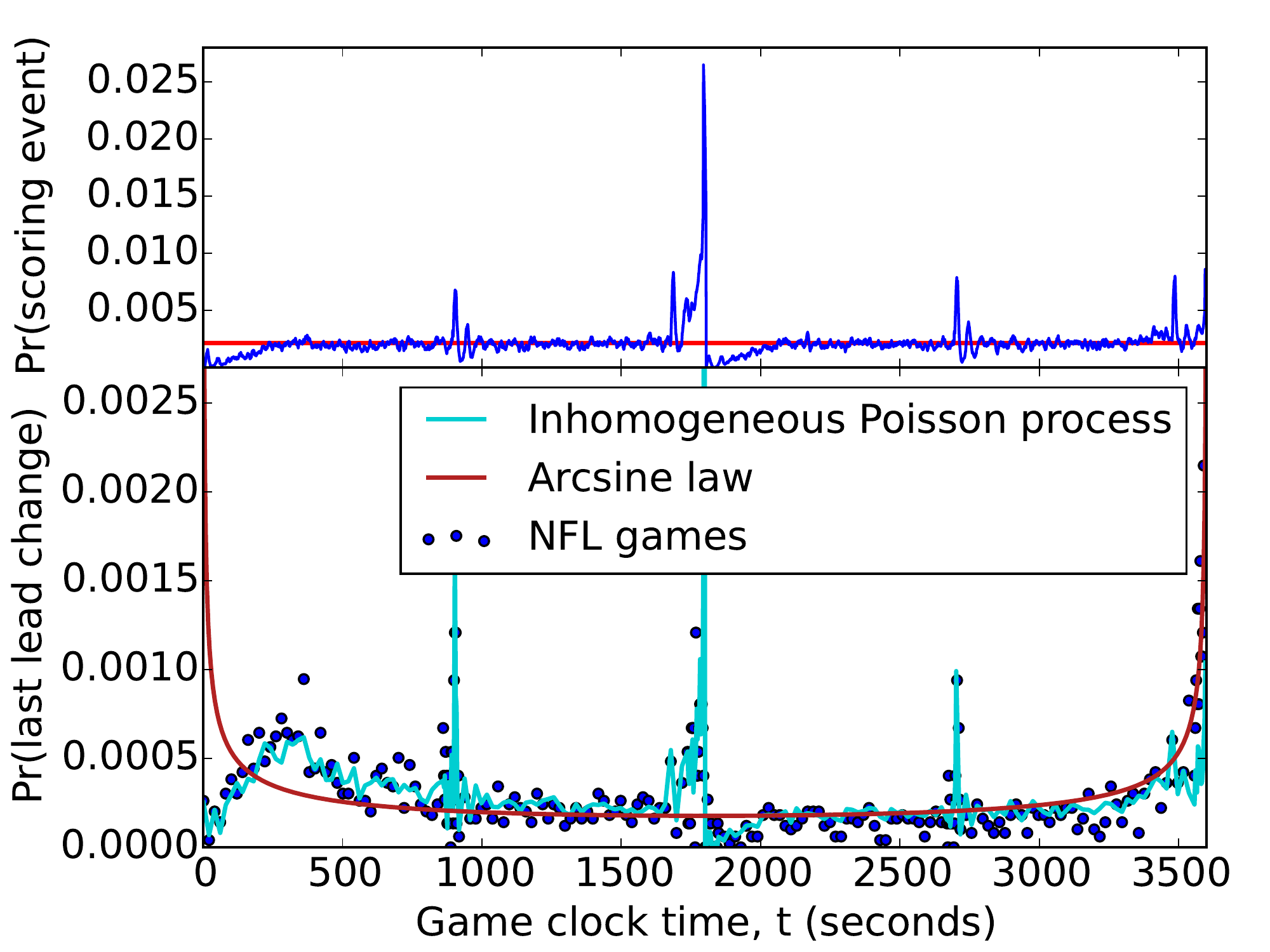} &
\includegraphics[scale=0.284]{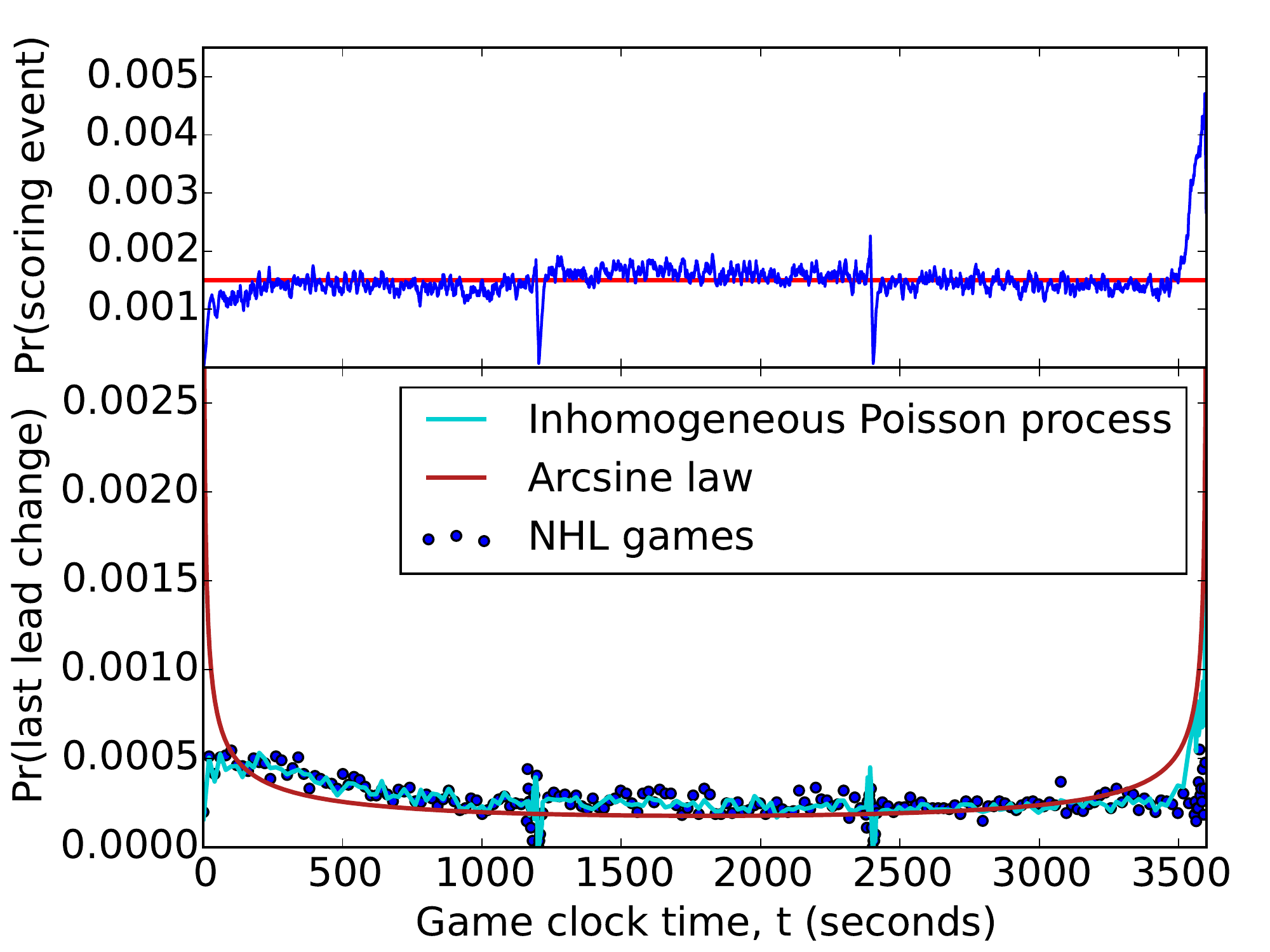}
\end{tabular}
\caption{(upper) Empirical probability that a scoring event occurs at time $t$, with the game-average scoring rate shown as a horizontal line, for games of CFB, NFL, and NHL. (lower) Distribution of times $\mathcal{L}(t)$ for the last lead change. }
\label{fig:last:lead:change}
\end{figure*}

\begin{figure*}[t!]
\begin{tabular}{ccc}
\includegraphics[scale=0.290]{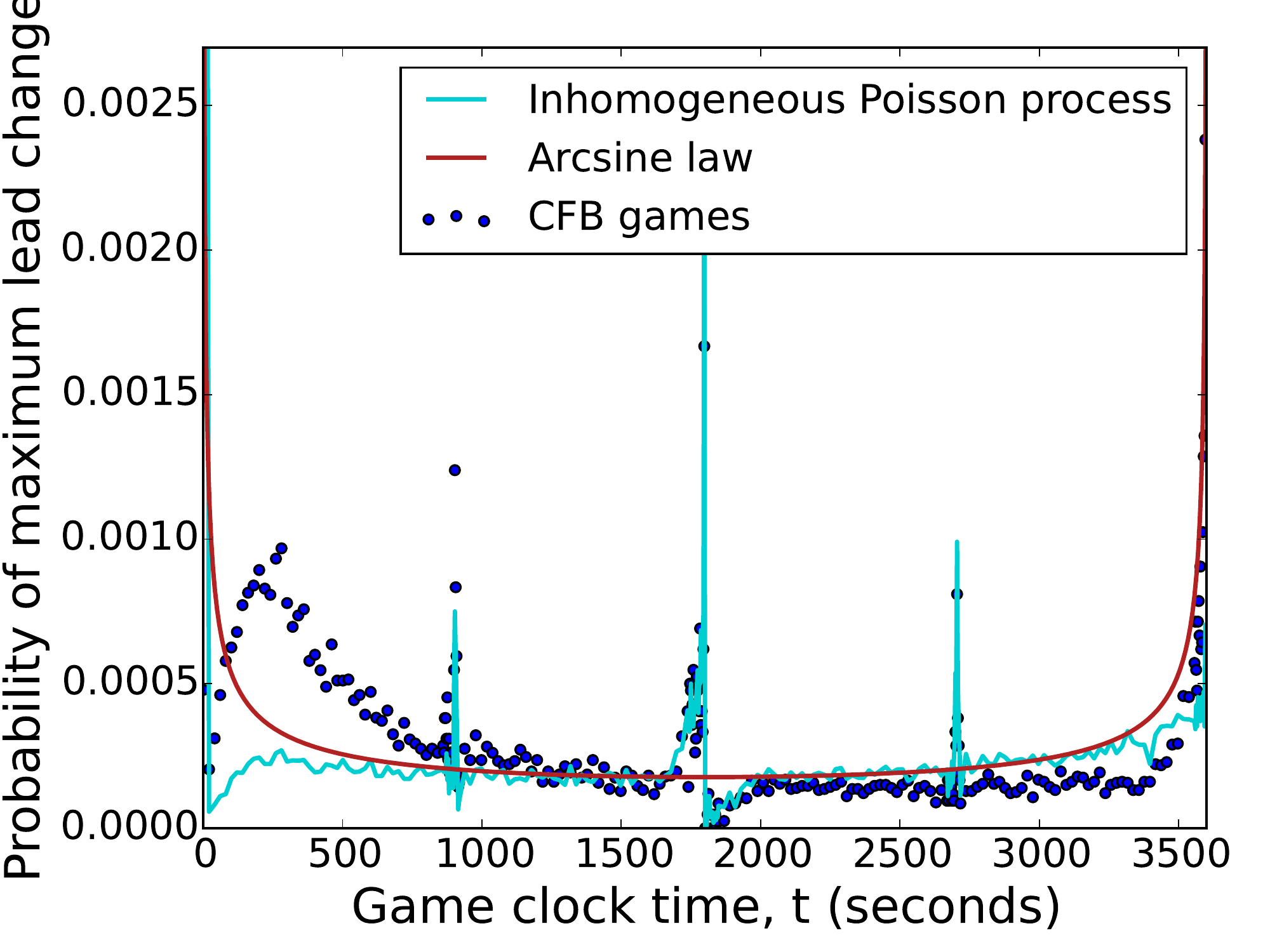} &
\includegraphics[scale=0.290]{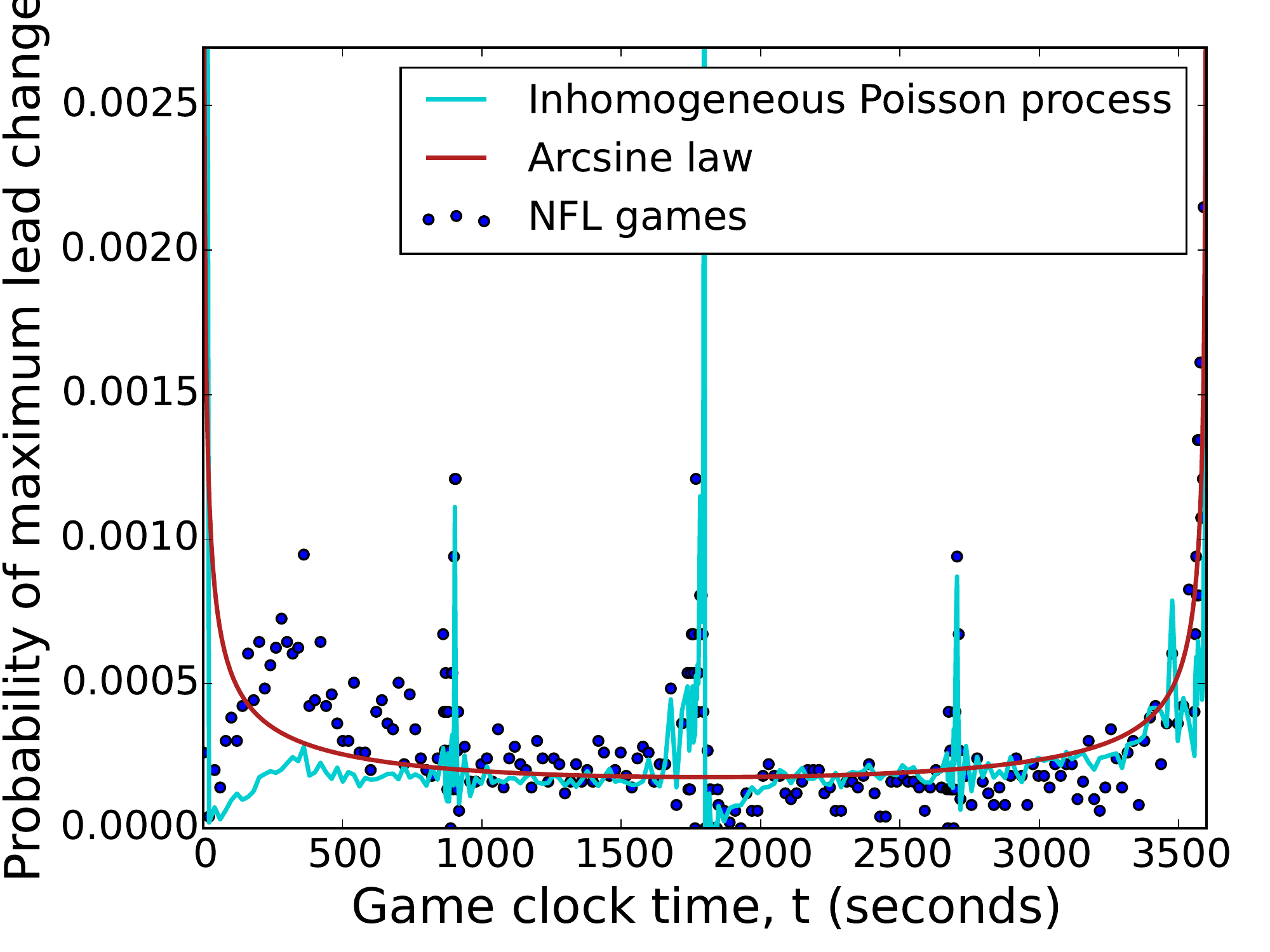} &
\includegraphics[scale=0.290]{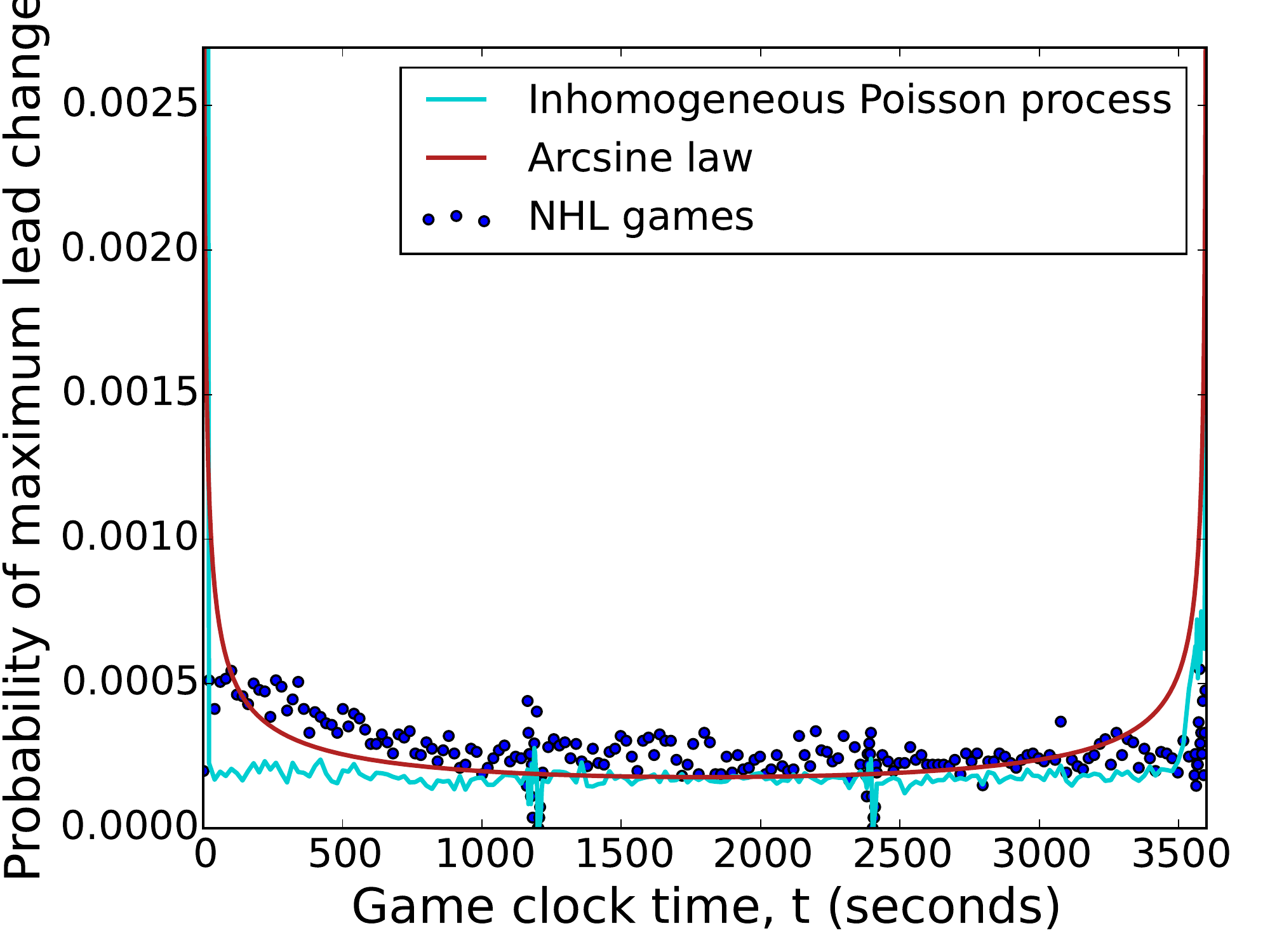}
\end{tabular}
\caption{Distribution of times $\mathcal{M}(t)$ for the maximal lead, for
  games of CFB, NFL, and NHL. }
\label{fig:max:lead}
\end{figure*}

However, there are also points of departure, the most important of which is
that the scoring rate in these sports is between 10--25 times smaller than in
basketball.  Because of this much lower overall scoring rate, the diminished
rate at the start of games is much more apparent than in basketball
(Fig.~\ref{fig:last:lead:change}).  This longer low-activity initial period
and other non-random-walk mechanisms cause the distributions $\mathcal{L}(t)$
and $\mathcal{M}(t)$ to visibly deviate from the arcsine laws
(Figs.~\ref{fig:last:lead:change} and \ref{fig:max:lead}).  A particularly
striking feature is that $\mathcal{L}(t)$ and $\mathcal{M}(t)$ approach zero
for $t\to 0$.  In contrast, because the initial reduced scoring rate occurs
only for the first 30 seconds in NBA games, there is a realu, but barely
discernible deviation of the data for $\mathcal{L}(t)$ from the arcsine law
(Fig.~\ref{Lt}).

\begin{figure*}[t!]
\begin{tabular}{ccc}
\includegraphics[scale=0.303]{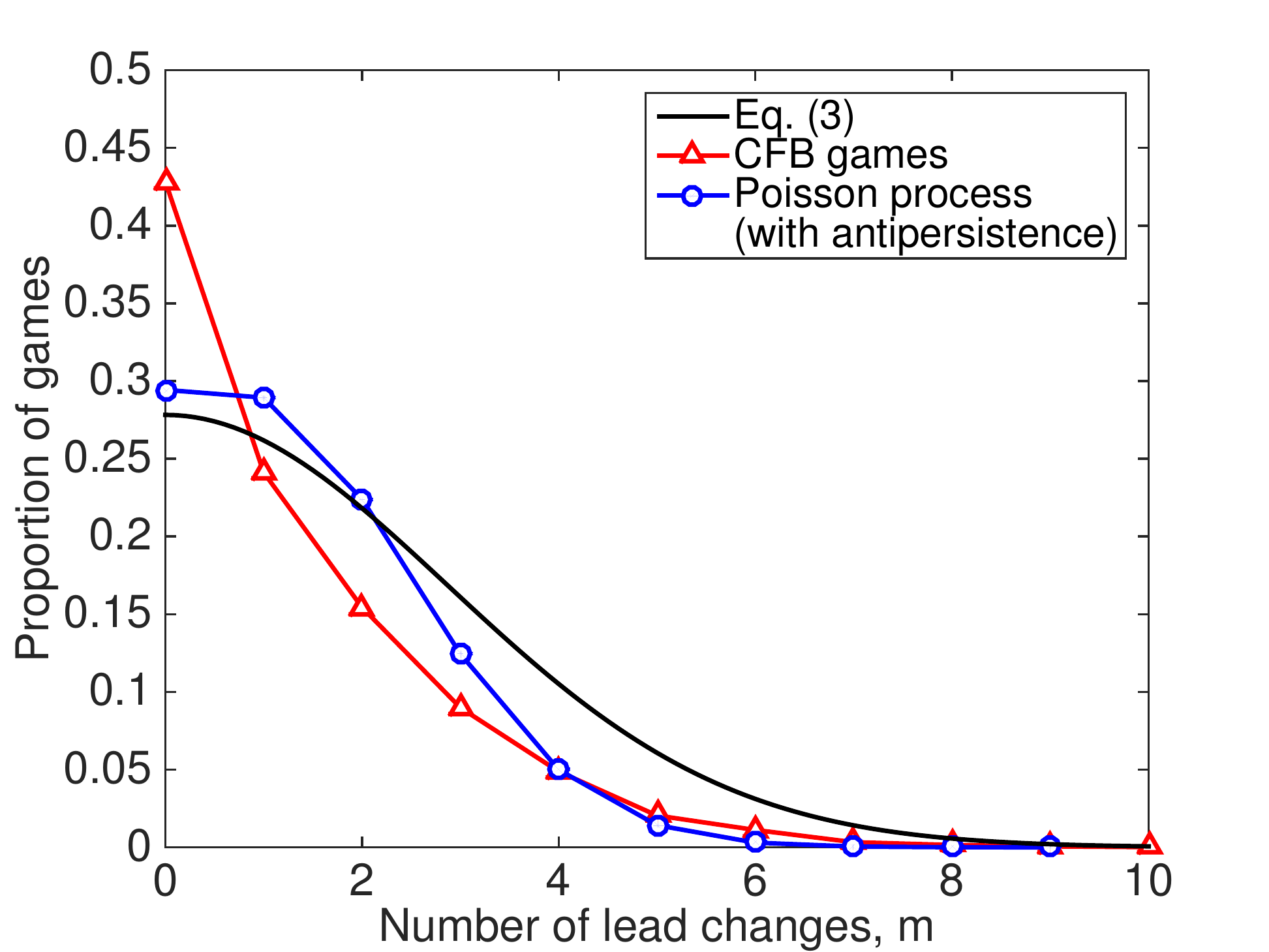} &
\includegraphics[scale=0.303]{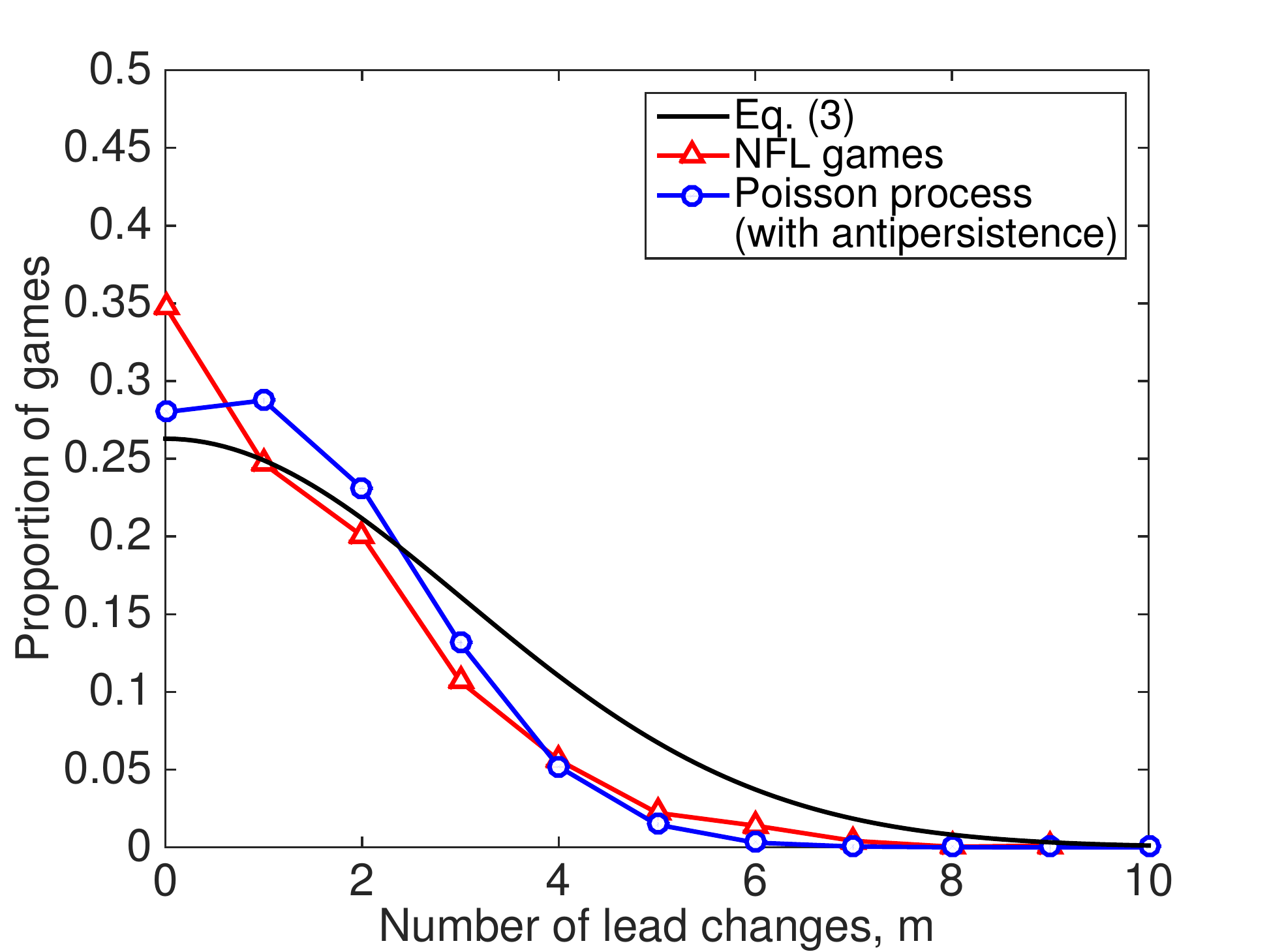} &
\includegraphics[scale=0.303]{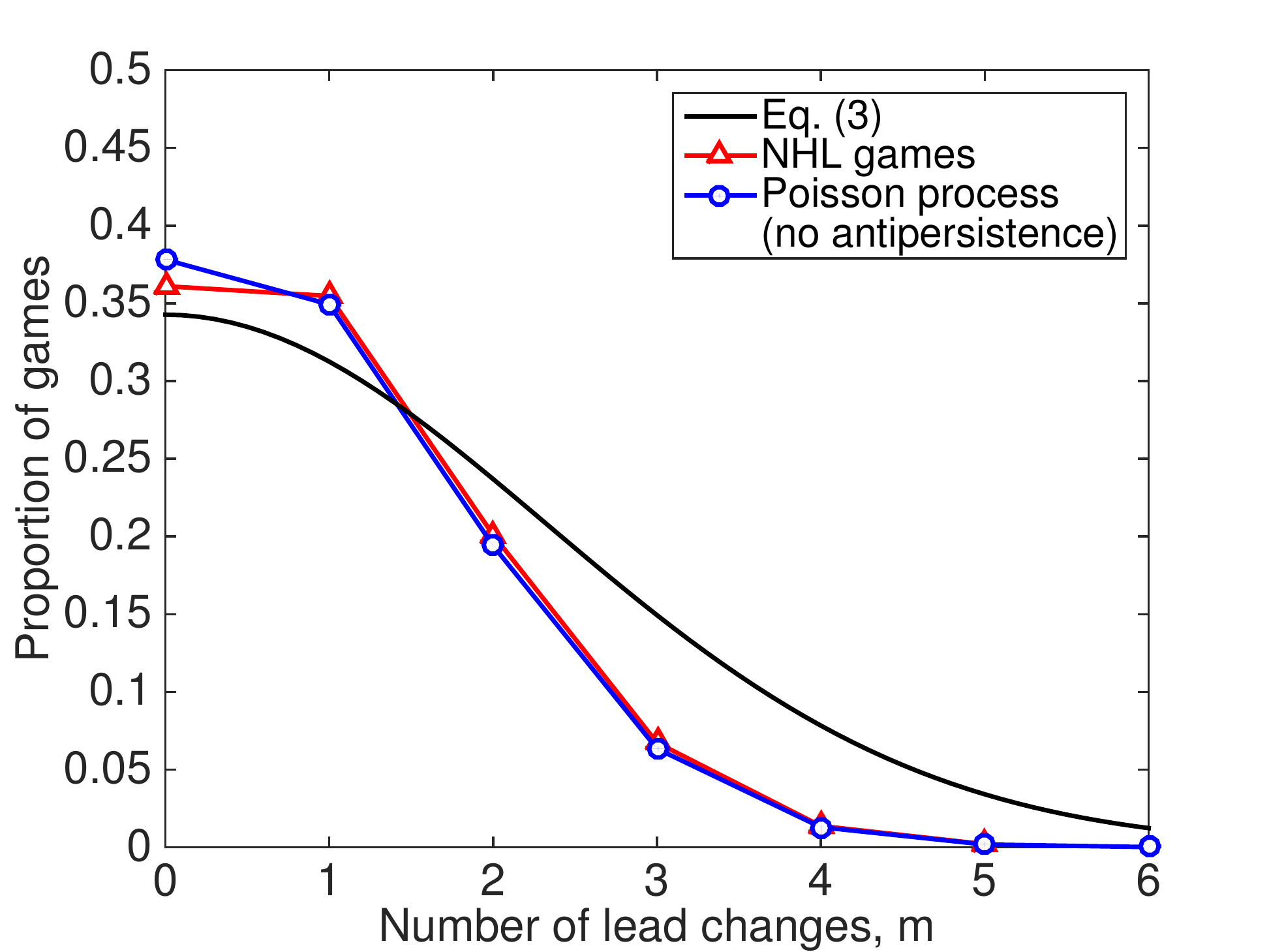}
\end{tabular}
\caption{Distribution of the average number of lead changes per game, for
  CFB, NFL, and NHL, showing the simple prediction of Eq.~\eqref{G}, the
  empirical data, and the results of a simulation in which scoring events
  occur by a Poisson process with the game-specific scoring rate.}
\label{fig:lead:changes:per:game}
\end{figure*}

\begin{figure*}[t!]
\begin{tabular}{ccc}
\includegraphics[scale=0.303]{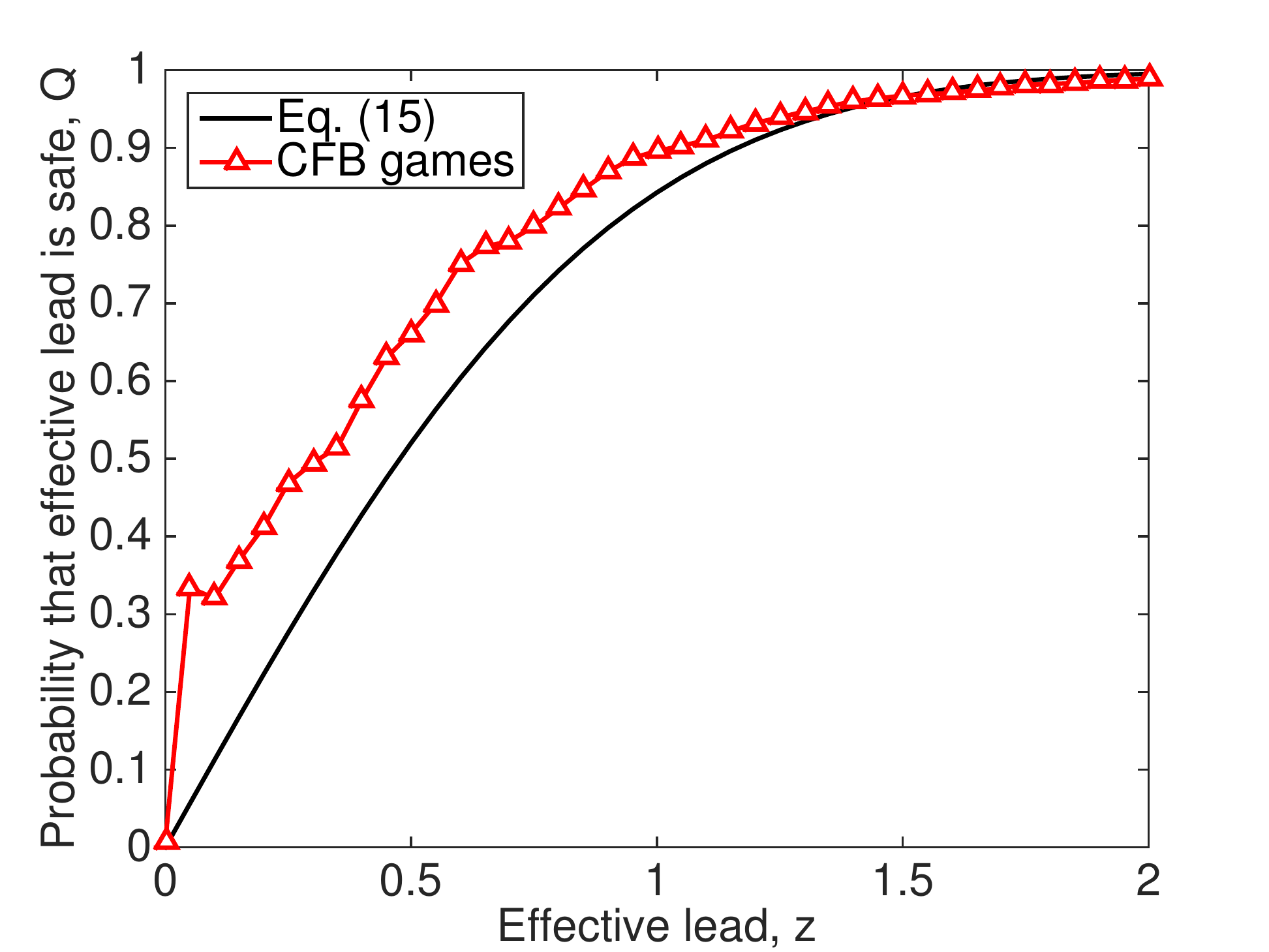} &
\includegraphics[scale=0.303]{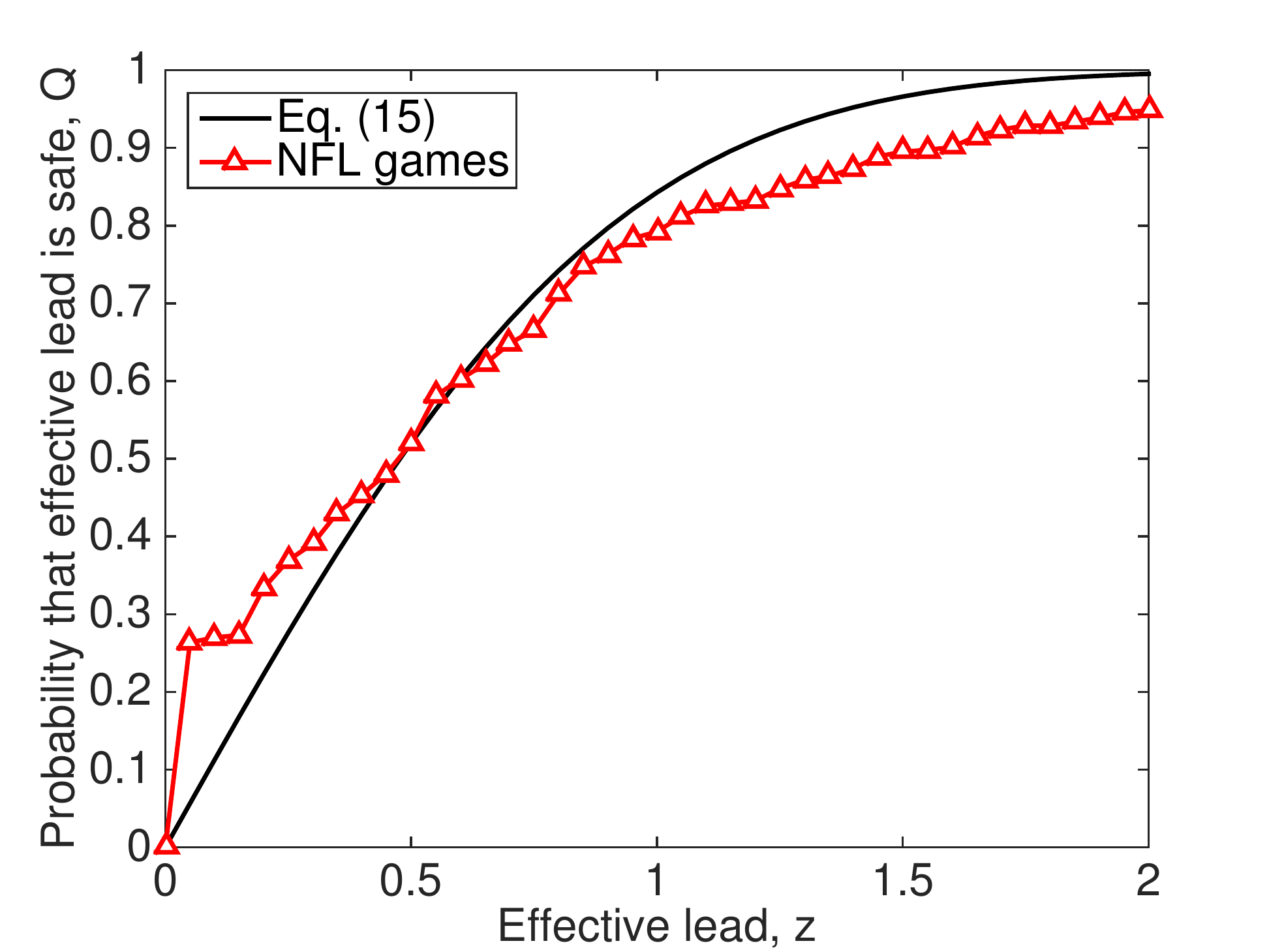} &
\includegraphics[scale=0.303]{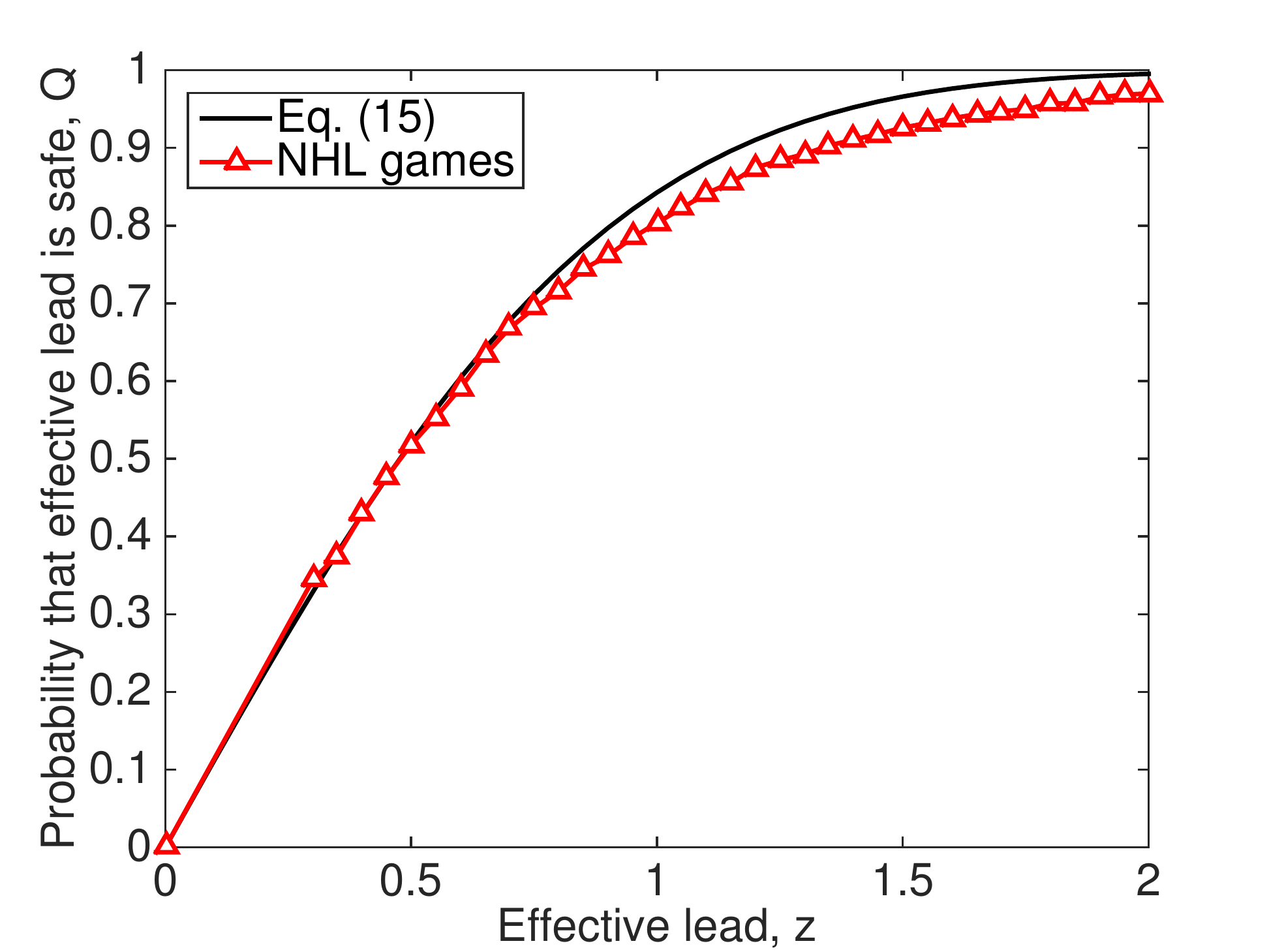}
\end{tabular}
\caption{Probability that a lead is safe, for CFB, NFL, and NHL, versus the
  dimensionless lead $z=L/\sqrt{4D\tau}$. Each figure shows the prediction
  from Eq.~\eqref{Q-final} and the corresponding empirical pattern. }
\label{fig:prob:safe}
\end{figure*}

Finally, the safe lead probability given in Eq.~\eqref{Q-final} qualitatively
matches the empirical data for football and hockey
(Fig.~\ref{fig:prob:safe}), with the hockey data being closest to the
theory~\footnote{For convenience, the 90\%-safe leads for CFB, NFL, and NHL
  are solutions to $L=\alpha\sqrt{\tau}$, where
  $\alpha=\{0.4629,0.3781,0.0638\}$, respectively.}.  For both basketball and
hockey, the expression for the safe lead probability given in
Eq.~\eqref{Q-final} is quantitatively accurate.  For football, a prominent
feature is that small leads are much more safe that what is predicted by our
theory.  This trend is particularly noticeable in the CFB.  One possible
explanation of this behavior is that in college football, there is a
relatively wide disparity in team strengths, even in the most competitive
college leagues.  Thus a small lead size can be quite safe if the
two teams happen to be significantly mismatched.

For American football and hockey, it would be useful to understand how the
particular structure of these sports would modify a random walk model.  For
instance, in American football, the two most common point values for scoring
plays are 7 (touchdown plus extra point) and 3 (field goal).  The random-walk
model averages these events, which will underestimate the likelihood that a
few high-value events could eliminate what otherwise seems like a safe lead.
Moreover, in football the ball is moved incrementally down the field through
a series of plays. The team with ball possession has four attempts to move
the ball a specific minimum distance (10 yards) or else lose possession; if
it succeeds, that team retains possession and repeats this effort to further
move the ball.  As a result, the spatial location of the ball on the field
likely plays an important role in determining both the probability of scoring
and the value of this event (field goal versus touchdown). In hockey, players
are frequently rotated on and off the ice so that a high intensity of play is
maintained throughout the game.  Thus the pattern of these substitutions---
between potential all-star players and less skilled ``grinders''---can change
the relative strength of the two teams every few minutes.

\section{Conclusions}

A model based on random walks provides a remarkably good description for the
dynamics of scoring in competitive team sports.  From this starting point, we
found that the celebrated arcsine law of Eq.~\eqref{arcsine} closely
describes the distribution of times for: (i) one team is leading
$\mathcal{O}(t)$ (first arcsine law), (ii) the last lead change in a game
$\mathcal{L}(t)$ (second arcsine law), and (iii) when the maximal lead in the
game occurs $\mathcal{M}(t)$ (third arcsine law).  Strikingly, these arcsine
distributions are bimodal, with peaks for extremal values of the underlying
variable.  Thus both the time of the last lead and the time of the maximal
lead are most likely to occur at the start or the end of a game.

These predictions are in accord with the empirically observed scoring
patterns within more than 40,000 games of professional basketball, American
football (college or professional), and professional hockey.  For basketball,
in particular, the agreement between the data and the theory is quite close.
All the sports also exhibit scoring anomalies at the end of each scoring
period, which arise from a much higher scoring rate around these
times (Figs.~\ref{Lt} and~\ref{fig:last:lead:change}).  For football and
hockey, there is also a substantial initial time range of reduced scoring
that is reflected in $\mathcal{L}(t)$ and $\mathcal{M}(t)$ both approaching
zero as $t\to 0$.  Football and hockey also exhibit other small but
systematic deviations from the second and third arcsine laws that remain
unexplained.

The implication for basketball, in particular, is that a typical game can be
effectively viewed as repeated coin-tossings, with each toss subject to the
features of antipersistence, an overall bias, and an effective restoring
force that tends to shrink leads over time (which reduces the likelihood of a
blowout). These features represent inconsequential departures from a pure
random-walk model.  Cynically, our results suggest that one should watch only
the first few and last few minutes of a professional basketball game; the rest of the
game is as predictable as watching repeated coin tossings.
On the other hand, the high degree of unpredictability of events in the
middle of a game may be precisely what makes these games so exciting for
sports fans.

The random-walk model also quantitatively predicts the probability that a
specified lead of size $L$ with $t$ seconds left in a game is ``safe,'' i.e.,
will not be reversed before the game ends.  Our predictions are
quantitatively accurate for basketball and hockey. For basketball, our
approach significantly outperforms a popular heuristic for determining when a
lead is safe.  For football, our prediction is marginally less accurate, and
we postulated a possible explanation for why this inaccuracy could arise in
college football, where the discrepancy between the random-walk model and the
data is the largest.

Traditional analyses of sports have primarily focused on the composition of
teams and the individual skill levels of the players.  Scoring events and
game outcomes are generally interpreted as evidence of skill differences
between opposing teams.  The random walk view that we formalize and test here
is not at odds with the more traditional skill-based view.  Our perspective
is that team competitions involve highly skilled and motivated players who
employ well-conceived strategies.  The overarching result of such keen
competition is to largely negate systematic advantages so that all that
remains is the residual stochastic element of the game.  The appearance of
the arcsine law, a celebrated result from the theory of random walks, in the
time that one team leads, the time of the last lead change, and the time at
which the maximal lead occurs, illustrates the power of the random-walk view
of competition.  Moreover, the random-walk model makes surprisingly accurate
predictions of whether a current lead is effectively safe, i.e., will not be
overturned before the game ends, a result that may be of practical interest
to sports enthusiasts.

The general agreement between the random-walk model for lead-change dynamics
across four different competitive team sports suggests that this paradigm has
much to offer for the general issue of understanding human competitive
dynamics.  Moreover, the discrepancies between the empirical data and our
predictions in sports other than basketball may help identify alternative
mechanisms for scoring dynamics that do not involve random walks.  Although
our treatment focused on team-level statistics, another interesting direction
for future work would be to focus on understanding how individual behaviors
within such social competitions aggregate up to produce a system that behaves
effectively like a simple random walk.  Exploring these and other hypotheses,
and developing more accurate models for scoring dynamics, are potential
fruitful directions for further work.

\begin{acknowledgements}
  The authors thank Sharad Goel and Sears Merritt for helpful
  conversations. Financial support for this research was provided in part by
  Grant No.\ DMR-1205797 (SR) and Grant No.\ AGS-1331490 (MK) from the NSF,
  and a grant from the James S.\ McDonnell Foundation (AC). By mutual agreement,
  author order was determined randomly, according to when the maximum lead
  size occurred during the Denver Nuggets--Milwaukee Bucks NBA game on 3 March 2015.
\end{acknowledgements}


%

\end{document}